\documentclass[useAMS,usenatbib]{mn2e}

\usepackage{epsfig}
\usepackage{graphicx}
\usepackage{amssymb}

\title{An estimate of the local ISW signal, and its impact on CMB anomalies}
\author[C. L. Francis and J. A. Peacock]
       {{Caroline L. Francis and John
	   A. Peacock\thanks{E-mail: jap@roe.ac.uk}} \\ 
SUPA\thanks{Scottish Universities Physics
	     Alliance} Institute for Astronomy, University of Edinburgh, Blackford
	 Hill, Edinburgh EH9 3HJ, UK. \\
}

\date{}

\pagerange{\pageref{firstpage}--\pageref{lastpage}}
\pubyear{2009}

\newcommand{\mpcoh}{\,\ensuremath{h^{-1}}\textrm{Mpc}}
\def\[{\begin{equation}}
\def\]{\end{equation}}
\def\gsim{\mathrel{\lower0.6ex\hbox{$\buildrel {\textstyle >}
 \over {\scriptstyle \sim}$}}}
\def\lsim{\mathrel{\lower0.6ex\hbox{$\buildrel {\textstyle <}
 \over {\scriptstyle \sim}$}}}
\def\deg{^\circ}

\begin{document}

\maketitle

\label{firstpage}

\begin{abstract}
\noindent 
We estimate the local density field in redshift shells to a
maximum redshift of $z=0.3$, using photometric redshifts for the
2MASS galaxy catalogue, matched to optical data from the SuperCOSMOS
galaxy catalogue. This density-field map is used to
predict the Integrated Sachs-Wolfe (ISW) CMB anisotropies that
originate within the volume at $z<0.3$. We investigate the impact of this
estimated ISW foreground signal on large-scale anomalies in the WMAP
CMB data. We find that removal of the foreground ISW signal from WMAP
data reduces the significance of a number of reported large-scale
anomalies in the CMB, including the low quadrupole power and the
apparent alignment between the CMB quadrupole and octopole.
\end{abstract}


\section{Introduction}

\label{sec:intro}

The possible presence of non-Gaussian features in the low-order
multipoles of the Cosmic Microwave Background has been a recurring
theme in cosmology for over a decade.
Large-scale anomalies of varying significance have been
claimed, including the low power in the CMB quadrupole
\citep{Hinshaw_1996,Spergel_lowl2}; the planarity of the octopole and
its apparent alignment with the quadrupole
\citep{Tegmark_l2l3align,deOliveira}; the north-south power asymmetry
about the ecliptic plane \citep[e.g.][]{Eriksen_northsouth} and the
extension of the $\ell=2$ and $\ell=3$ alignments to higher multipoles
\citep{Schwarz_2004,Copi_quad+oct},
defining the `Axis of Evil' \citep{Land_Magueijo_2005}. A closely related
result is the lack of large-angle correlations in the CMB sky
\citep{Hinshaw_1996,Copi_2009}.
On smaller scales, \citet{Vielva_coldspot} suggest that the CMB contains a
non-Gaussian `cold spot' in the southern hemisphere. There are four
possible explanations for these CMB anomalies: (i) 
a problem with the data; (ii) an unlucky fluke;
(iii) new early-universe physics; or (iv) they are due to a
foreground signal. Here we explore the last of these
possibilities.

The ISW effect is a secondary anisotropy in the CMB caused by the
passage of CMB photons through evolving gravitational potential wells:
\[
\frac{\Delta T^{\textrm{\tiny ISW}}}{T_{\textrm{\tiny CMB}}} =
2\int_{t_{\textrm{\tiny
LS}}}^{t_{0}}{\frac{\dot{\Phi}(\vec{x}(t),t)}{c^{2}}} \, \mathrm{d}t
\label{eq:isw},
\]
where $t_{0}$ and $t_{\textrm{\tiny LS}}$ denote the times
today and at last scattering respectively; $\vec{x}$ is the position
along the line of sight of the photon at time $t$ and $\Phi$ is the
gravitational potential \citep{Martinez-Gonzalez_Sanz_Silk}. On large
scales in a Dark Energy dominated universe, CMB photons gain energy
when they pass through the decaying potential wells associated with
overdensities and lose energy on passing through
underdensities. This leads to the
large-scale positive cross-correlation between large scale structure
and the CMB that is often used to attempt detections of the ISW effect
\citep{Crittenden_Turok_1996}.

The ISW effect is the only significant secondary CMB anisotropy at low
$\ell$. The only other non-negligible signal, due to the thermal SZ
effect, is approximately an order of magnitude smaller than the
expected ISW signal. Our aim here is to predict the local
ISW signal to a maximum redshift of $z=0.3$ and consider the effect of
removing this part of the foreground signal on the large-scale CMB
anomalies. The data and method used to predict the local ISW signal
are described in Section \ref{sec:data} and the effect of this
foreground signal on large-scale CMB anomalies in explored in Section
\ref{sec:results}. Finally, our conclusions are presented in Section
\ref{sec:concl}. A cosmological model with $\Omega_m = 0.3$,
$\Omega_v=0.7$, $\Omega_b=0.05$, $h=0.7$ and $\sigma_{8}=0.75$ is
assumed unless otherwise stated.

%%%%%%%%%%%%%%%%%%%%%%%%%%%%%%%%%%%%%%%%%%%%%%%%%%%%%%%%%%%%%%%%%%%%%%%%%%%%

\section{Prediction of the local ISW signal}

\label{sec:data}

In order to predict the local ISW signal, it is first necessary to
estimate the density field within the volume of interest. We make a 2D
Wiener reconstruction of the projected density field in redshift
shells, using photometric redshift data for the 2MASS galaxy
catalogue. An account of the data and method used for this
reconstruction is given in Section \ref{sec:data_LSS} and the
estimation of the local ISW signal from this density field is
described in Section \ref{sec:data_ISW}.

\subsection{Recovery of the local density field \label{sec:data_LSS}}

\subsubsection{Data}

The 2MASS dataset used in this paper is described briefly below. This
material is common to a companion paper about attempted detection of
the ISW effect via cross-correlation \citep{Francis_ISW} and fuller
details are given there.

2MASS is an all-sky survey in the $J$, $H$ and $K_{s}$ bands
\citep{Jarrett_2MASS}. This near-infrared selection means that 2MASS is
sensitive to old stars and hence to the most massive structures, which
makes it well-suited for constructing maps of galaxy mass density. The
final extended source catalogue (XSC) contains over 1.6 million
objects, over $98\%$ of which are galaxies. Photometric redshifts for
the 2MASS XSC have been generated by matching the 2MASS data with
optical catalogues from SuperCOSMOS scans of major photographic sky
surveys (UKST in the south; POSS2 in the north) to create a 5-band
\emph{BRJHK} dataset. Details of the SuperCOSMOS catalogue
construction process are given in \citet{Hambly_SuperCOSMOS} and
details of the photometric calibration and photo-$z$ determination
method are given in \citet{Peacock_photoz}. The final photometric
redshift dataset has an overall rms in $z_{\mathrm{phot}}-z$ of
0.033.

A magnitude cut $K_{s}<13.8$ is imposed on the extinction-corrected
data to ensure uniformity across the region outside the Galactic
plane. The Galactic plane mask is constructed using the dust maps of
\citet{Schlegel_dust} and excludes all regions with $K$-band reddening
$A_{K}>0.05$. This leaves approximately $67\%$ of the sky
unmasked. Whilst the choice of $A_{K}>0.05$ is conservative,
\citet{Afshordi} note that the number density within pixels of faint
galaxies ($13.5<K_{20}<14.0$) drops off as a function of $A_{K}$ for
$A_{K}>0.065$. A less conservative mask, for example with $A_{K}>0.1$,
would still only leave $79\%$ of the sky unmasked -- at the price of
possible incompleteness issues in the analysis.

\subsubsection{Density field reconstruction}

It is possible to attempt a full 3D reconstruction of the local
density field with this photometric redshift dataset, and then use
this to estimate the local ISW signal (see
\citealt{Francis_PHD}). However, this recovered density field is
subject to significant radial smearing due to the uncertainty in the
radial position of each galaxy (approximately $100\mpcoh$).  The
extent of this smearing justifies a simpler and more transparent
approach: to recover the projected density field for 3 thick redshift
slices of width $\Delta z=0.1$ to $z_{\mathrm{max}}=0.3$, and estimate
the ISW signal due to each slice.

The galaxy distribution in each photometric redshift shell is known
outside the masked region. We use a Wiener filtering technique to
recover the galaxy density field in each shell. When
the Wiener filter is applied to all-sky data, it recovers the optimal
density field in the least squares sense; a thorough discussion of 
Wiener-filter density reconstruction is presented in
e.g. \citet{Zaroubi_1995}.

\begin{figure*}
\begin{minipage}[l]{0.48\linewidth}
\epsfig{file=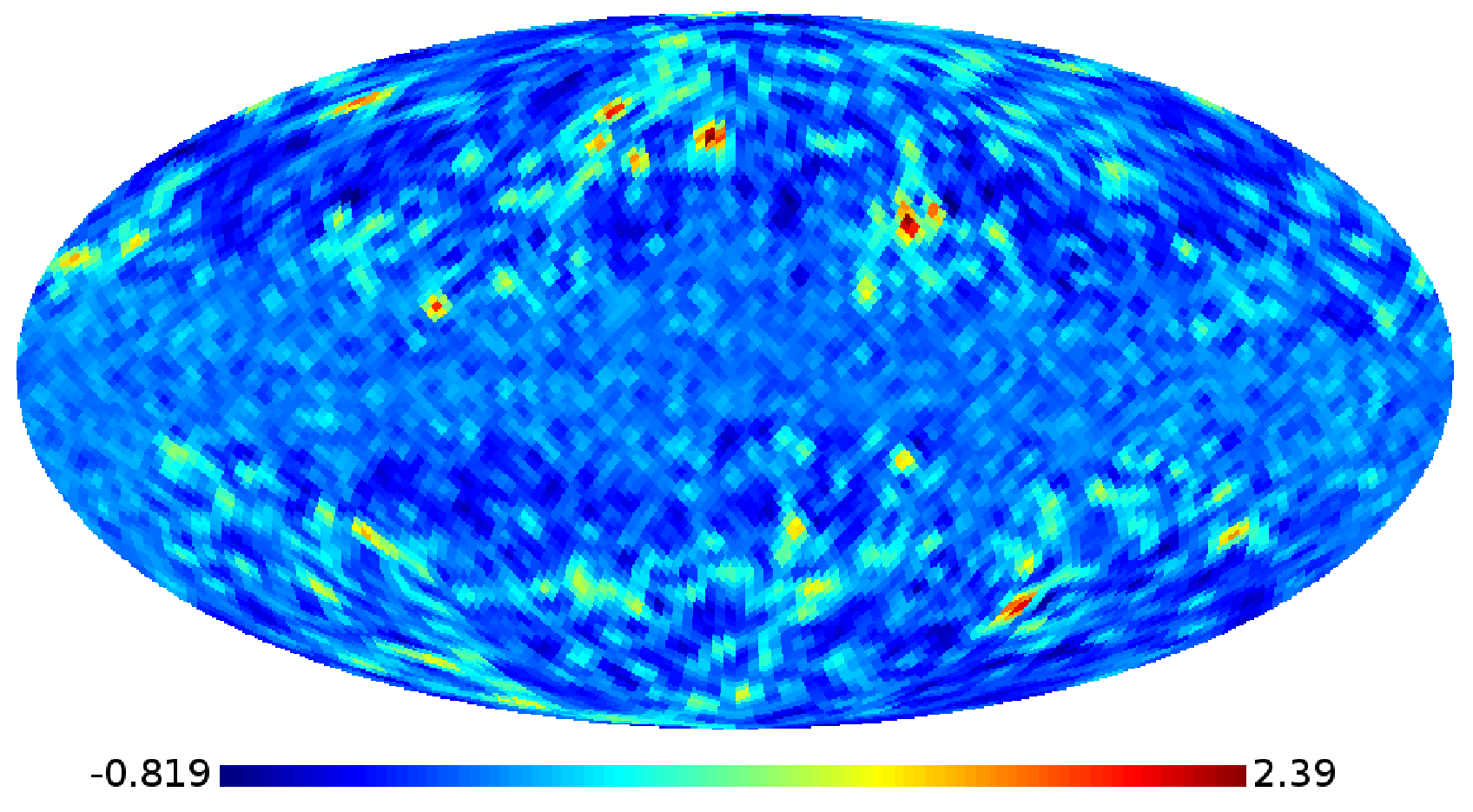,width=\linewidth,height=5cm}
\end{minipage}
\hspace{0.5cm}
\begin{minipage}[r]{0.48\linewidth}
\epsfig{file=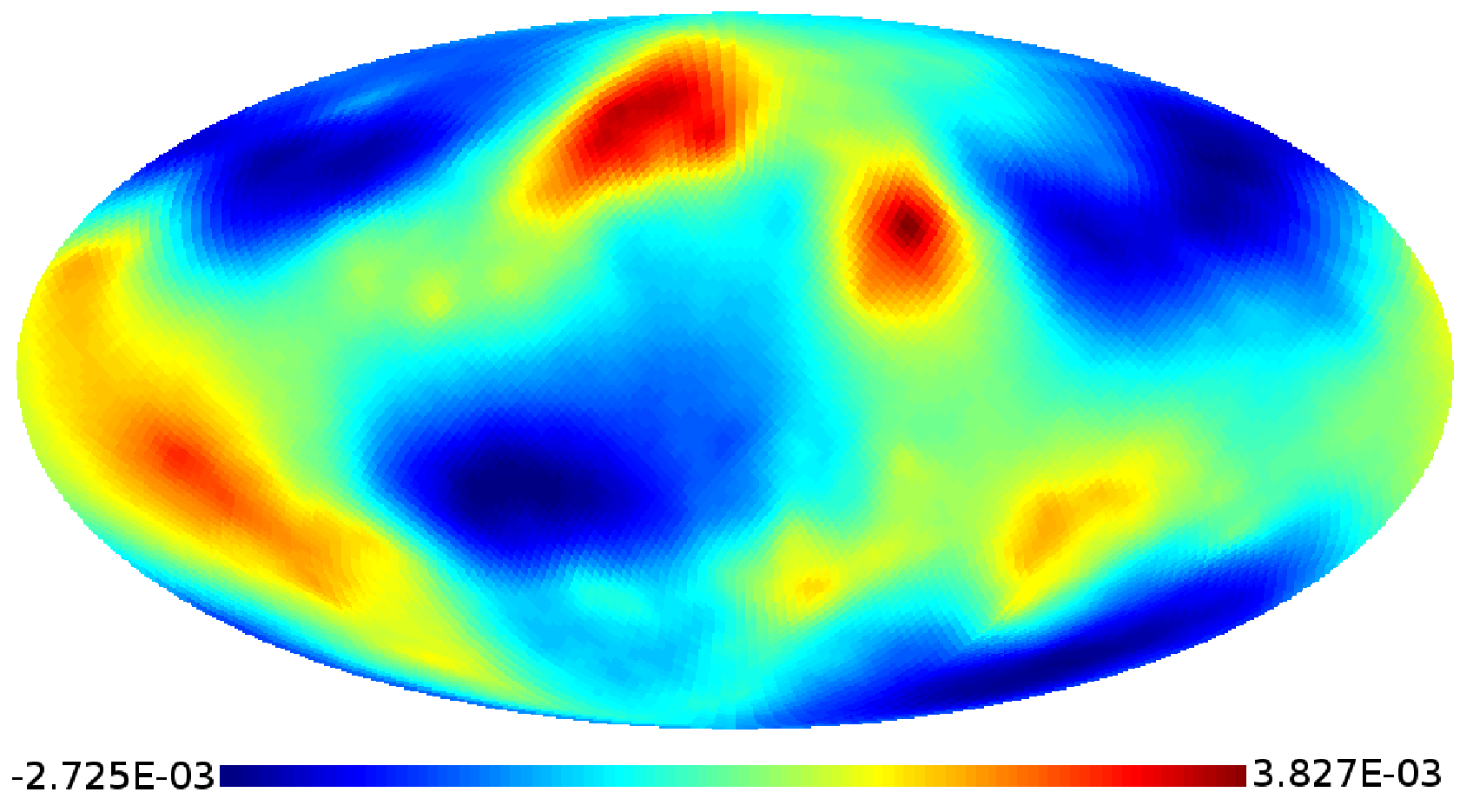,width=\linewidth,height=5cm}
\end{minipage}
\begin{minipage}[l]{0.48\linewidth}
\epsfig{file=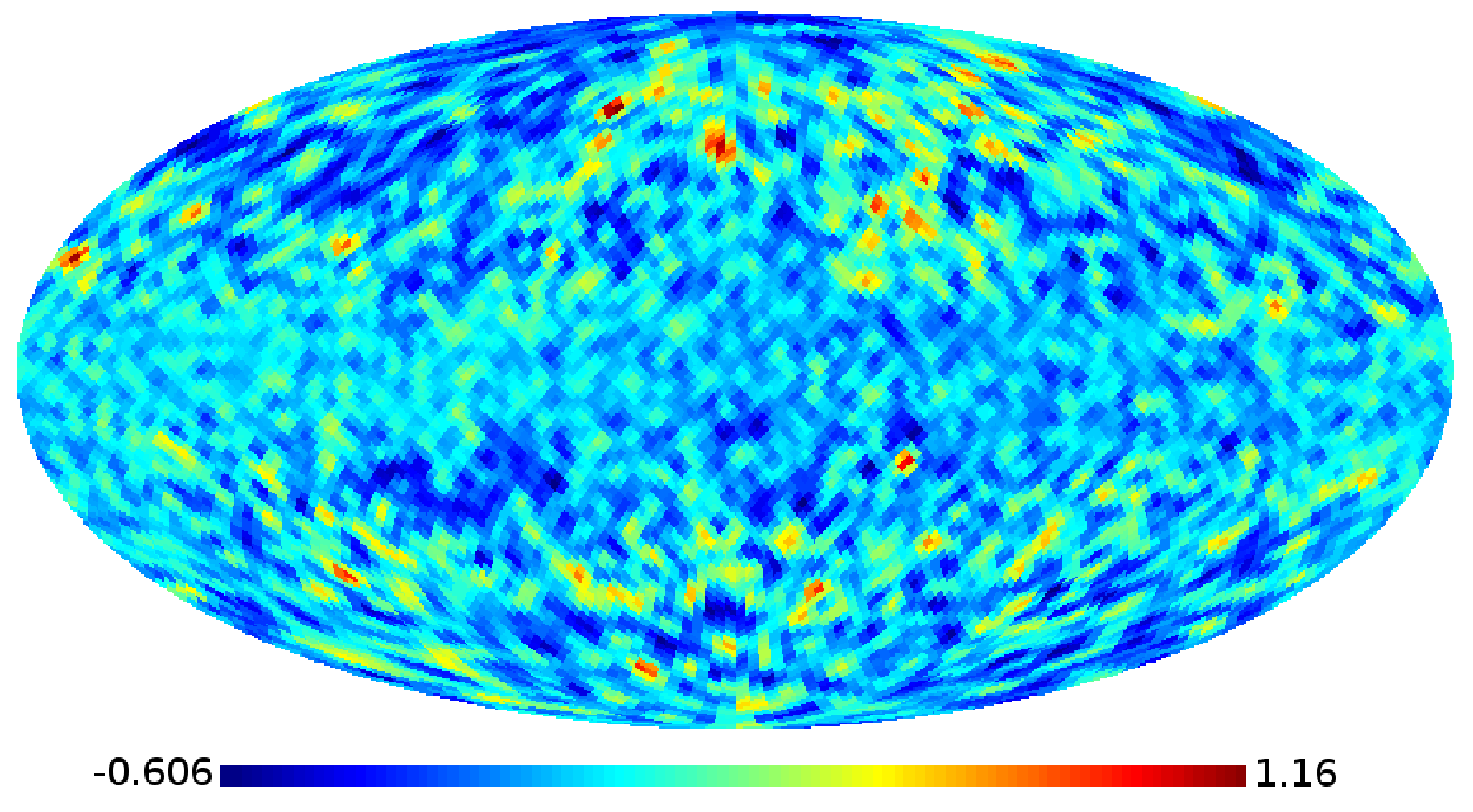,width=\linewidth,height=5cm}
\end{minipage}
\hspace{0.5cm}
\begin{minipage}[r]{0.48\linewidth}
\epsfig{file=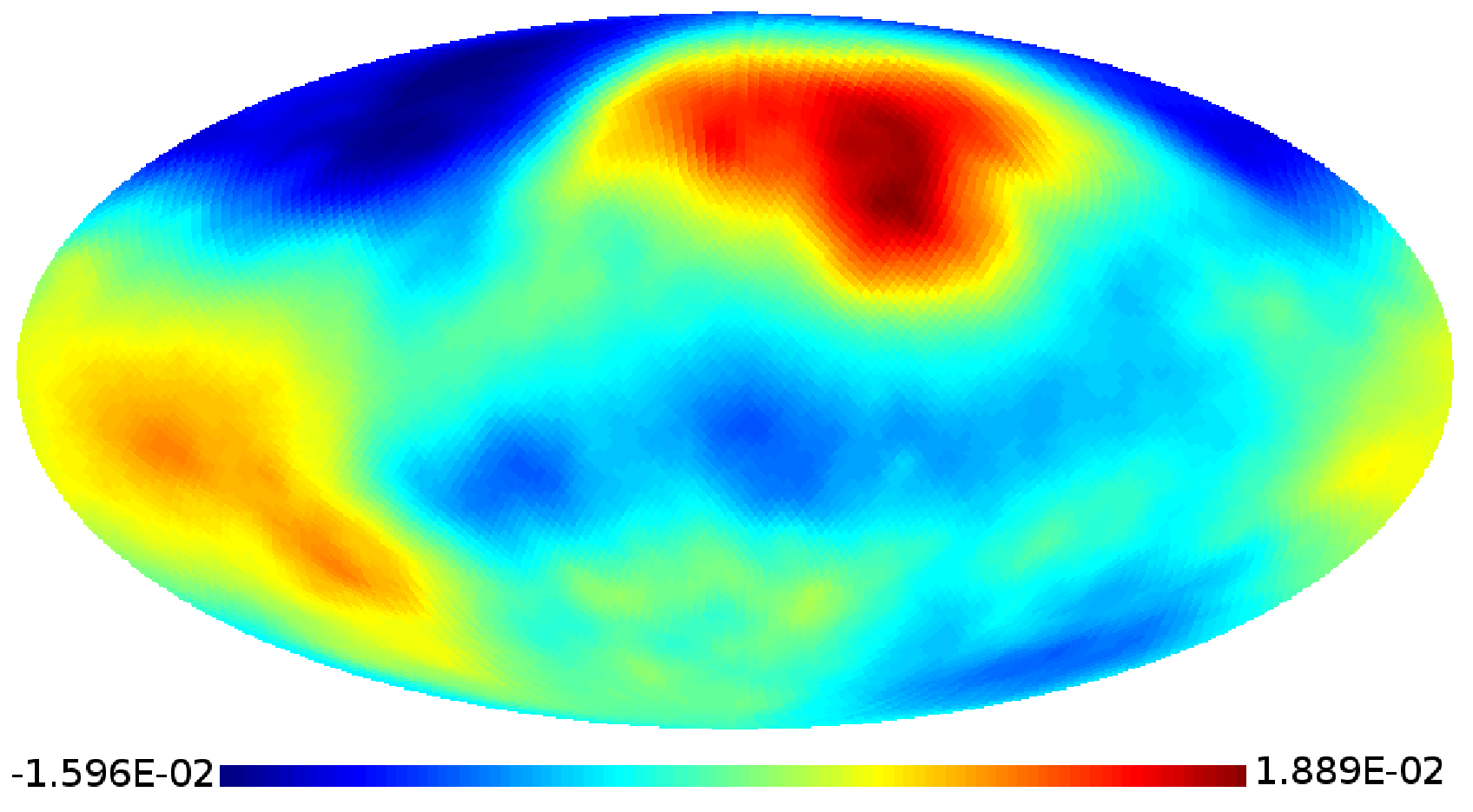,width=\linewidth,height=5cm}
\end{minipage}
\begin{minipage}[l]{0.48\linewidth}
\epsfig{file=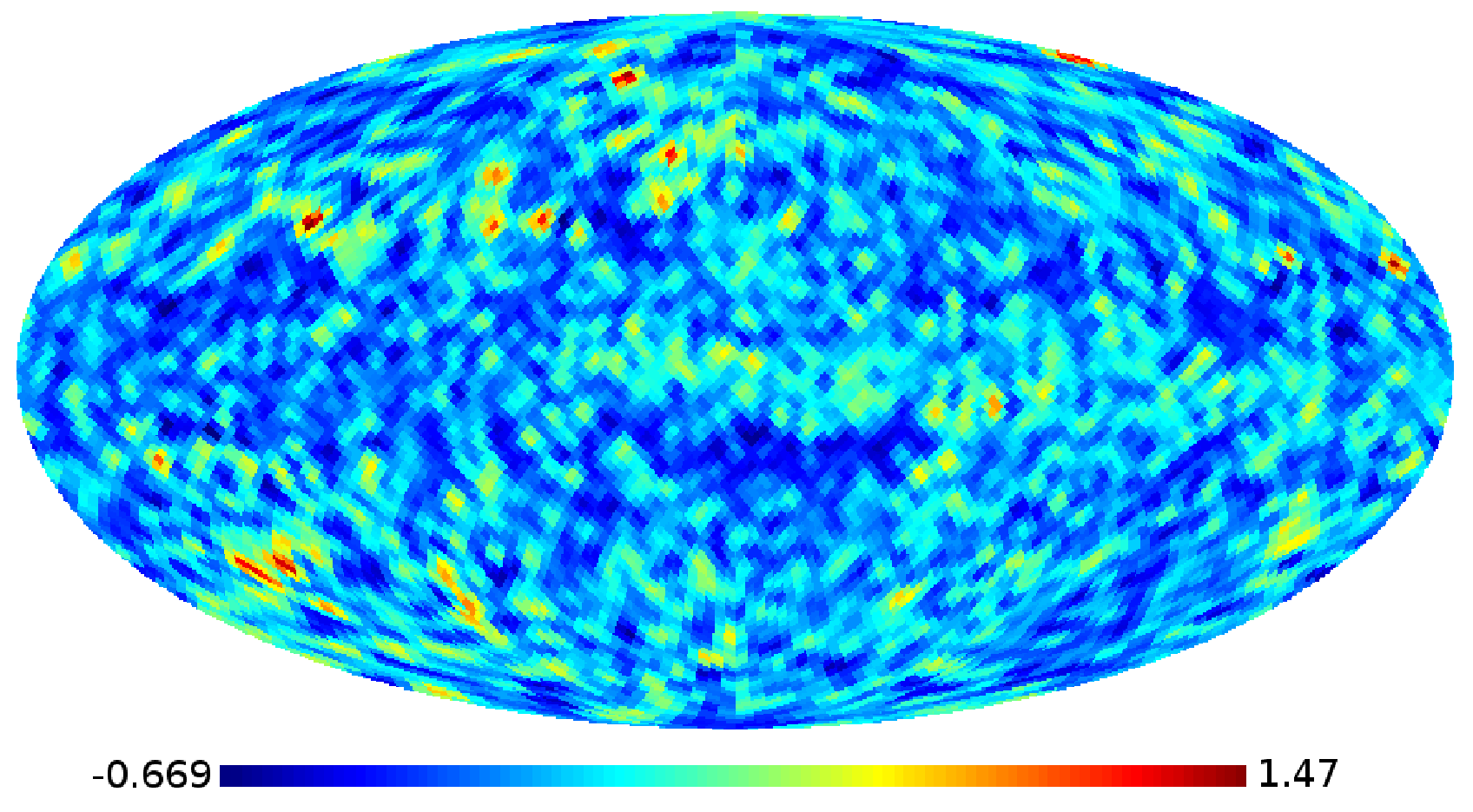,width=\linewidth,height=5cm}
\end{minipage}
\hspace{0.5cm}
\begin{minipage}[r]{0.48\linewidth}
\epsfig{file=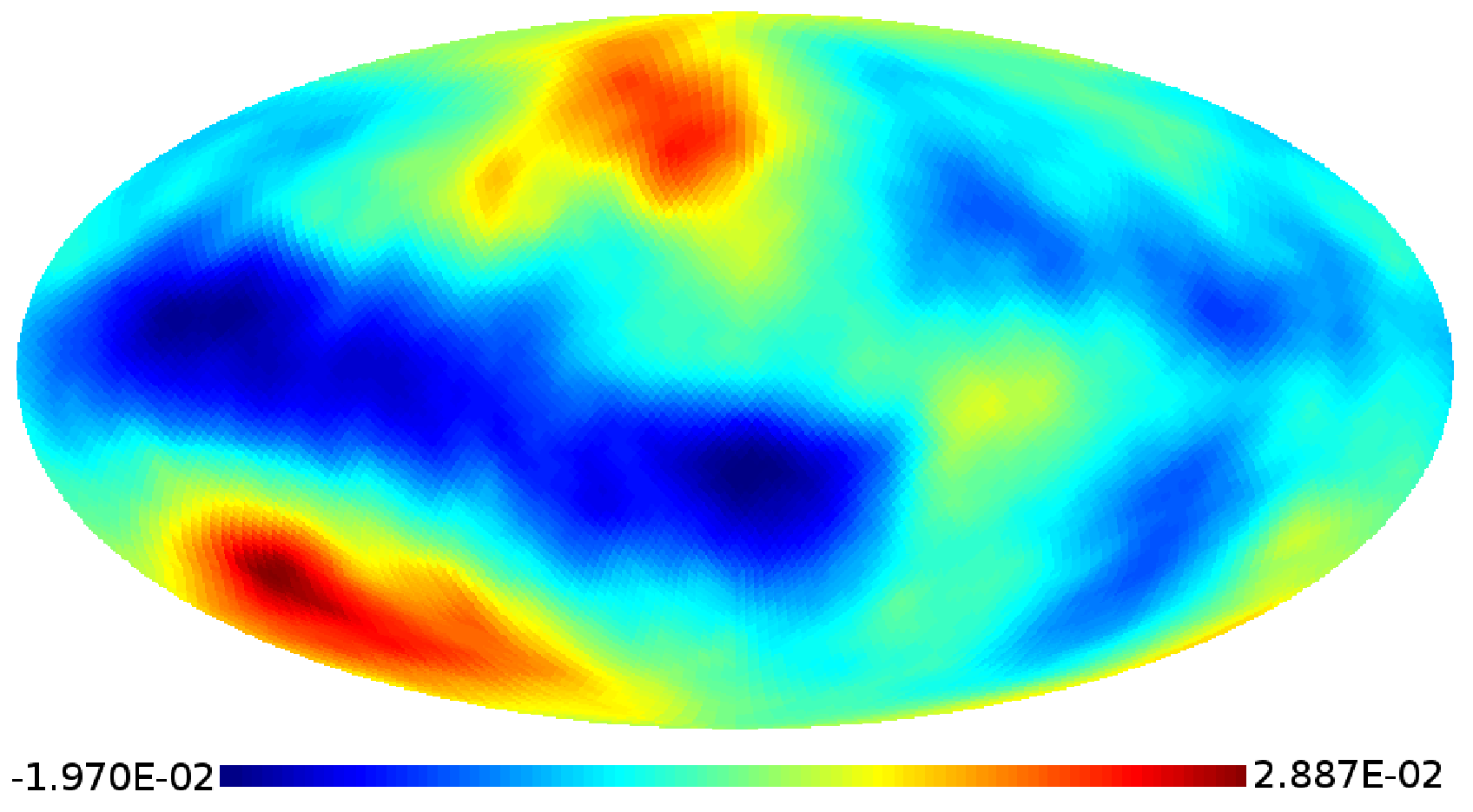,width=\linewidth,height=5cm}
\end{minipage}
\caption{(Left) The 2D reconstruction of the local density field
  described in Section \ref{sec:data_LSS} in three photometric
  redshift shells: $0.0<z<0.1$ (top), $0.1<z<0.2$ (middle) and
  $0.2<z<0.3$ (bottom). The plots show overdensity $\delta$ on a scale
  $-0.6 \le \delta \le 0.6$. (Right) The corresponding ISW signal in
  mK computed from the reconstructed density field using equation
  (\ref{eq:iswmap}). \label{fig:ISW_from_gals}}
\end{figure*}

Because our dataset is masked at low Galactic latitudes, the simple
all-sky Wiener approach must be modified.  We proceed by initially
filling the masked region in each of the redshift shells with a random
Poisson sampling of galaxies according to the average number density
observed outside the plane. This galaxy distribution is then used to
deduce the projected galaxy density field for the redshift shell. The
Wiener filter appropriate for each shell is then constructed as usual:
\[
\Phi_{\mathrm{WF}}(\ell) =
\frac{C_{\mathrm{gg}}(\ell)}{C_{\mathrm{gg}}(\ell)+C_{\mathrm{noise}}(\ell)},
\]
where $C_{\mathrm{gg}}(\ell)$ is the angular power spectrum of
galaxies in that redshift shell and $C_{\mathrm{noise}}(\ell)$
describes shot noise. The Wiener filter is applied in the first
instance to the $a_{\ell m}$ coefficients computed for the expansion
of this initial full-sky galaxy density field into a basis of
spherical harmonics. The resultant density field within the masked
region is Poisson sampled for galaxies once more to yield a revised
galaxy distribution composed of a sample from the Wiener
reconstruction within the plane, and the original 2MASS galaxies in
the unmasked regions.  In this way, large-scale features in surface
density can be extrapolated across the plane in a consistent way.  The
process can be iterated, but this did not yield a significant
improvement in the final results.  We have carried out tests of this
method on simulated data, and find satisfactory recovery of the
estimated ISW signal (Francis 2008).  Fig. \ref{fig:ISW_from_gals}
shows the recovered projected density field for each of the three
shells.
We use HEALPix software (Gorski et al. 2005; see also
\texttt{http://healpix.jpl.nasa.gov}) to generate these maps
and to perform harmonic analysis of them.

\subsection{Prediction of the local ISW signal \label{sec:data_ISW}}

Given a projected galaxy density field in a thick redshift shell, it
is possible to estimate the CMB temperature fluctuations, $\Delta T$, 
due to the ISW effect. Poisson's equation
tells us the time variation of the potential field in 
terms of the density growth factor in the linear regime, $g(a)$:
\[
\Phi(a) = \frac{g(a)}{a}\,\Phi(a=1).
\]
Hence, using Poisson's equation, we can integrate through a shell of thickness $\Delta r$:
\begin{eqnarray}
\frac{\Delta T_{\ell m}}{T} &\!\!\!=\!\!\!& -2\int
\frac{d}{dt}\left[\frac{g(a)}{a}\right]{\frac{a^{2}\Phi_{\ell
m}(a)}{g(a)}} \, \frac{dr}{c^{3}} \nonumber \\ &\!\!\!\simeq\!\!\!&
\frac{-2}{c^{3}}H(\bar{a})
\left(\frac{dg}{da}(\bar{a})-\frac{g(\bar{a})}{\bar{a}}\right)\frac{\bar{a}^{2}}{g(\bar{a})}
\int{\Phi_{\ell m}} \, dr \nonumber \\ 
&\!\!\!\simeq\!\!\!&
\frac{3H_{0}^{2}\Omega_{m}}{\ell(\ell+1)c^{3}}\left(1 -
\bar{a}\frac{g'(\bar{a})}{g(\bar{a})}\right)r^{2}(\bar{a})H(\bar{a})
\frac{\delta_{\ell m}}{b}\Delta r, \label{eq:iswmap}
\end{eqnarray}
where $\delta$ is the projected galaxy density field in the redshift
shell under consideration and $\bar{a} = (1+\bar{z})^{-1}$ with
$\bar{z}$ the redshift at the midpoint of the shell. The bias in each
redshift shell between the galaxy and matter density fields is
approximated using a linear bias relation $\delta_{g}=b\delta_{m}$ and
is assumed to be independent of scale and redshift in each shell. To
deduce the appropriate value of $b$ in each redshift shell, a maximum
likelihood fit of the predicted galaxy angular power spectrum to the
measured galaxy angular power spectrum is performed as in
\citet{Francis_ISW}.

\begin{figure*}
\centering
\begin{minipage}[l]{0.49\linewidth}
\begin{center}
\epsfig{file=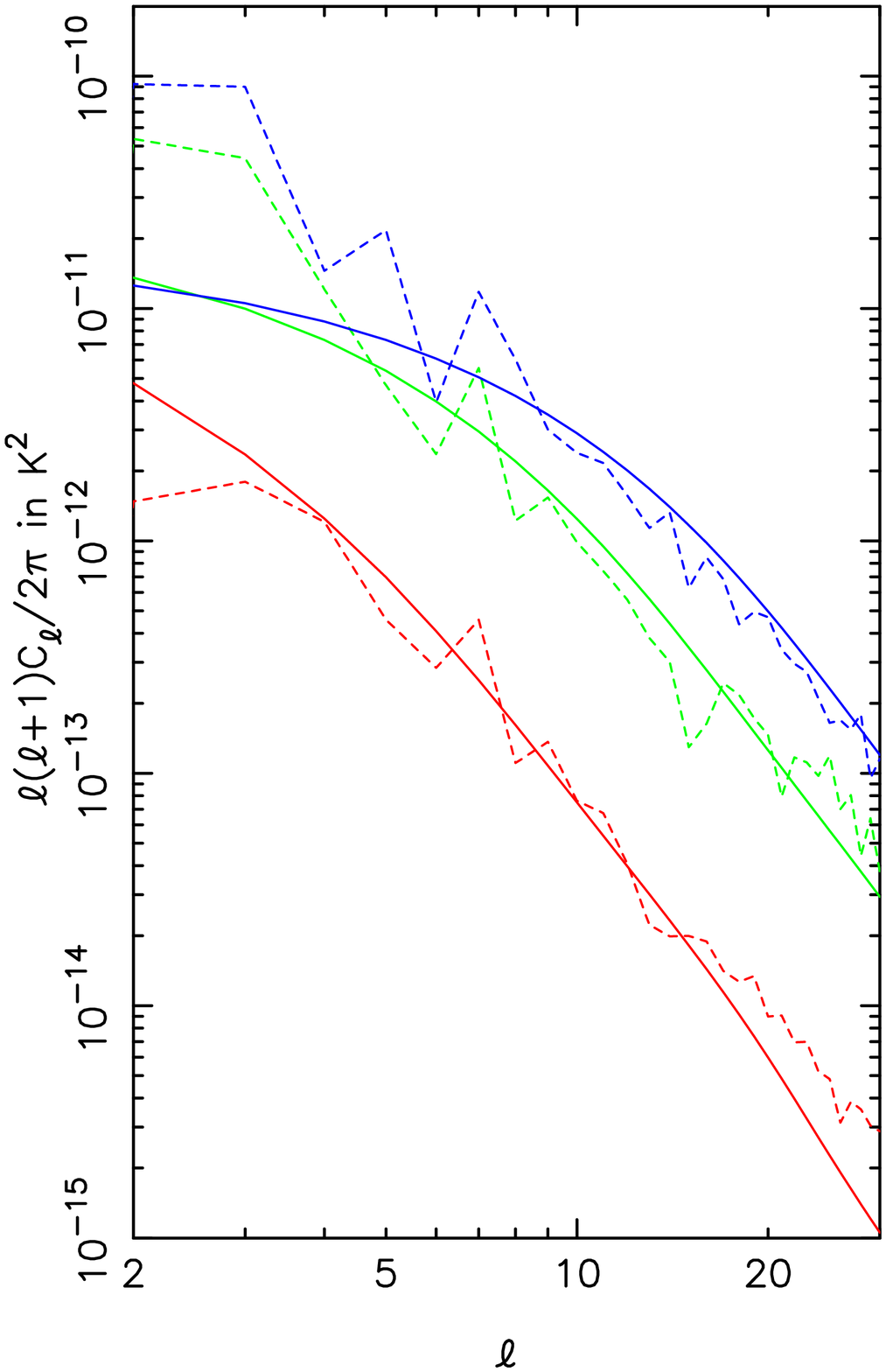,height=0.3\textheight,width=0.7\linewidth}
\end{center}
\end{minipage}
\begin{minipage}[r]{0.49\linewidth}
\begin{center}
\epsfig{file=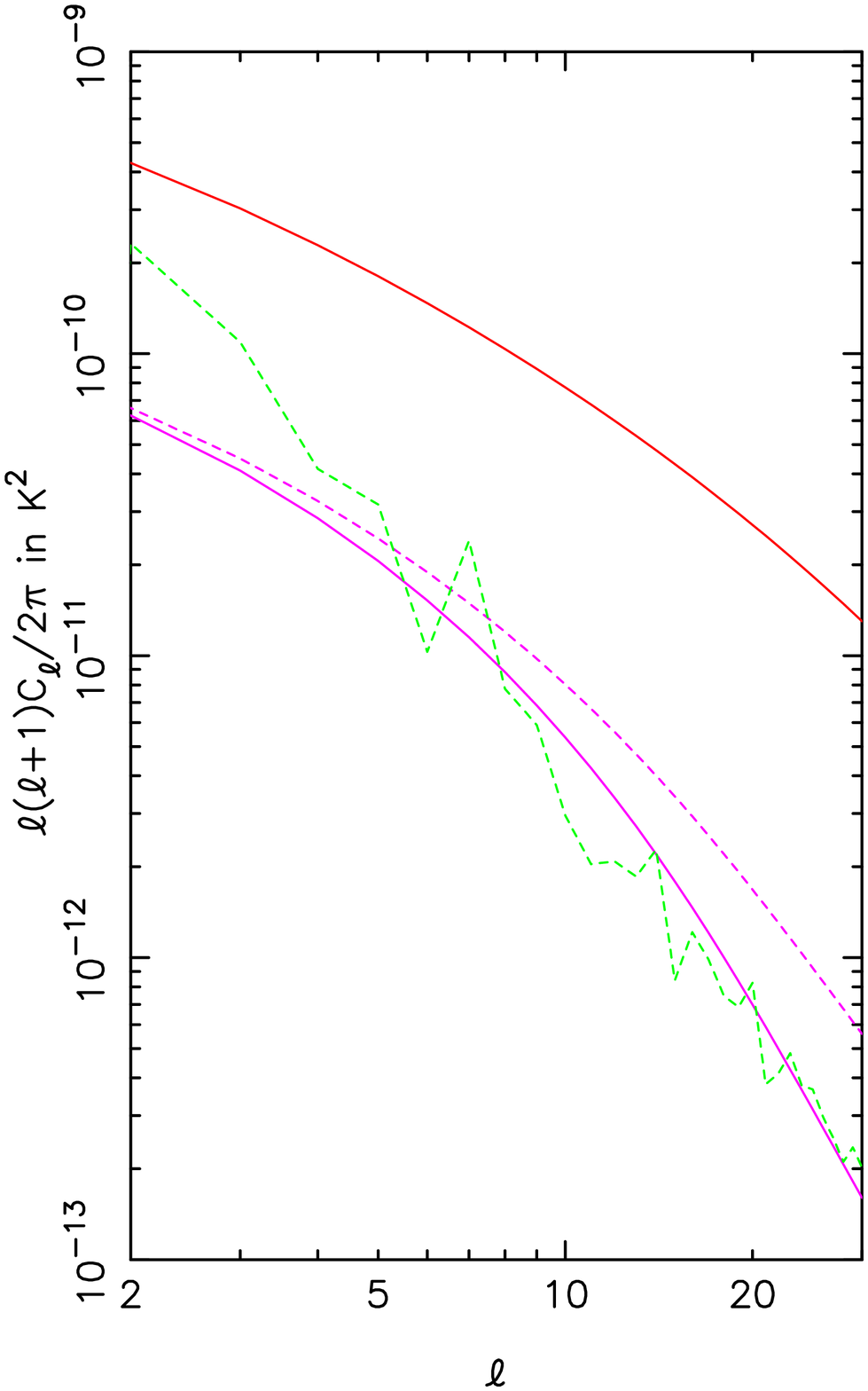,height=0.3\textheight,width=0.7\linewidth}
\end{center}
\end{minipage}
\caption{(\emph{Left}) The power spectrum of the ISW signal estimated
  in each of the three redshift shells (dashed lines): $0.0<z<0.1$
  (bottom), $0.1<z<0.2$ (middle) and $0.2<z<0.3$ (top) together with
  the predicted signal for $\Omega_m=0.3$ (solid lines), allowing for
  radial smearing. (\emph{Right}) The predicted total ISW power for
  $z_{\mathrm{max}}\rightarrow\infty$ (upper red solid line) and for
  $z_{\mathrm{max}}=0.3$ (middle purple dashed line) and that expected for
  $z_{\mathrm{max}}=0.3$ allowing for radial smearing (lower purple solid
  line). The green dashed line shows the power spectrum measured for
  the estimated $z_{\mathrm{max}}=0.3$ ISW
  signal. \label{fig:ISW_angpow}}
\end{figure*}

Fig. \ref{fig:ISW_from_gals} shows the results of estimating the local
ISW effect from the Wiener filtered density field in redshift
shells. 
These ISW maps are affected by
overlap effects due to the uncertainty in the radial
position of each galaxy: galaxies associated with an overdensity near
the boundary of a shell will to some extent be spread across both the
`true' shell and the adjacent redshift
shell. Fig. \ref{fig:ISW_angpow} shows the angular power spectrum of
the ISW signal in each slice, together with the predicted linear ISW
signal. For these predictions, we have accounted for the smearing
effects of the photometric redshift data by assuming that photo-$z$'s
apply Gaussian smoothing along the radial axis with
$\sigma_{r}=90\mpcoh$.  Fig. \ref{fig:ISW_angpow} also shows the
effect of this radial smearing on the total predicted ISW signal to
$z_{\mathrm{max}}=0.3$.  Overall, the agreement between observed and
predicted power spectra is good, which illustrates that the effect of
photo-$z$ imperfection is relatively minor for $\ell\lsim 20$.

We note that the estimated ISW signal in the lower two redshift slices
is slightly larger than expected at $\ell\gsim20$; this is particularly
noticeable in the $z<0.1$ slice. The nonlinear
Rees-Sciama effect is expected to increase the power at
high $\ell$, doubling the power at $\ell\simeq 200$
\citep[e.g.][]{Cooray_and_Sheth2002}. Since the mean depth of
the lowest-redshift shell is of order one tenth of the distances
that dominate the total ISW effect, it is plausible that we are
seeing some local Rees-Sciama effect. In any case, the main focus of
the present paper is at larger angular scales.
We also note that the estimated ISW power
is larger than expected for $\ell\lsim 3$ in the $0.1<z<0.2$ shell and
for $\ell\lsim5$ in the $0.2<z<0.3$ shell, which leads to a larger
than average total signal for $z_{\mathrm{max}}=0.3$ on such
scales. Such small-$\ell$ modes are potentially sensitive to the
corrections for mask incompleteness near the plane, but tests of our
method on mock data show no tendency to bias the power high in this
regime (although it does increase the already considerable cosmic
variance).  Any discrepancy is relative to our standard
model of $\Omega_m=0.3$, which is perhaps on the
high-density side given current data \citep{KomatsuWMAP09}. Decreasing
$\Omega_m$ to 0.25 with other parameters held fixed raises the
expected quadrupole power for $z_{\mathrm{max}}=0.3$ by a factor 1.5;
our adopted normalization of $\sigma_8=0.75$ might also be increased,
so that the predicted low-$\ell$ ISW power could plausibly increase in
amplitude by a factor 2.  We will thus provisionally adopt the
large-scale ISW map as estimated, but will also explore the
consequences of scaling its amplitude down to the lower values
expected in our $\Omega_m=0.3$ model.

Fig. \ref{fig:ISW_angpow} also shows the total ISW signal for
$z_{\mathrm{max}}\rightarrow\infty$, which we can use to estimate the
fraction of the total ISW signal that we expect to include if
$z_{\mathrm{max}}=0.3$. The square root of the ratio of the predicted
ISW power for $z_{\mathrm{max}}\rightarrow\infty$ to that for
$z_{\mathrm{max}}=0.3$ shows that we can expect to include $\sim40\%$
of the total rms ISW temperature signal at low multipoles, falling to
$\sim10\%$ at $\ell=30$. This variation with angular scale is mainly due to
the greater importance of the high redshift ISW signal at large $\ell$.

%%%%%%%%%%%%%%%%%%%%%%%%%%%%%%%%%%%%%%%%%%%%%%%%%%%%%%%%%%%%%%%%%%%%%%%%%%%%%%%%%%%%

\section{Implications for CMB anomalies}

\label{sec:results}

Having made a prediction of the local ISW signal, we now investigate
the effect of this foreground signal on a number of the large scale
anomalies that have been reported in the CMB. Many of these CMB
anomalies relate solely to the $\ell = 2$ and $\ell = 3$ multipoles
and, as noted in Section \ref{sec:data_ISW}, the estimated local ISW
power is greater than the ensemble average on these scales. We
therefore consider two versions of the estimated local ISW signal in
this analysis of the CMB anomalies: the ISW prediction described in
Section \ref{sec:data} and a rescaled version of this signal chosen to
maintain the phases of the original signal but to have the power
spectrum expected of the ISW signal out to $z=0.3$ for our standard
choice of $\Omega_m=0.3$. Consideration of this latter ISW signal
allows us to assess the possibility that any effect on the CMB
anomalies is solely due to the large low-multipole power of the
estimated ISW signal.

\begin{figure*}
\begin{minipage}[l]{0.48\linewidth}
\epsfig{file=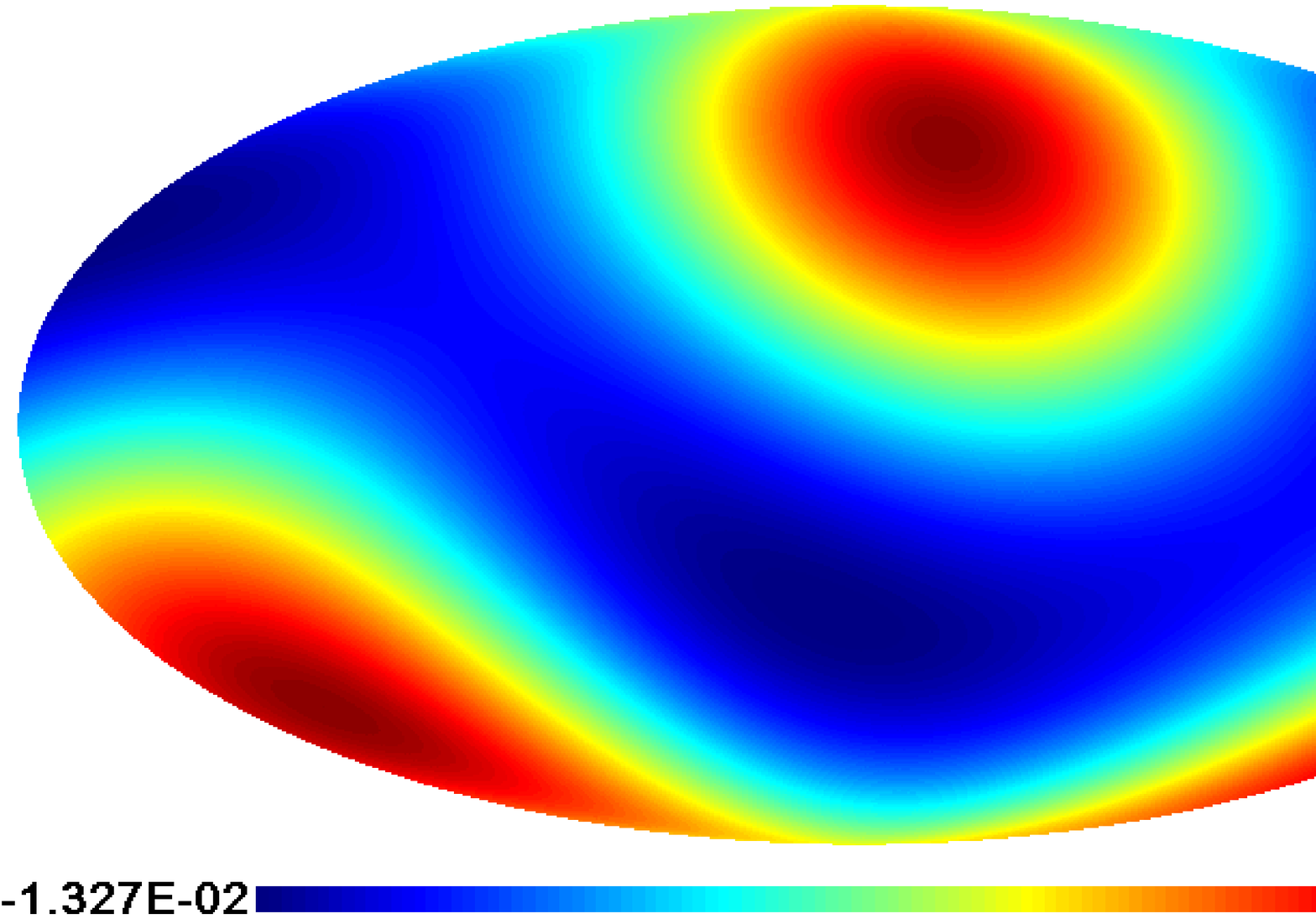,height=0.2\textheight,width=\linewidth}
\end{minipage}
\begin{minipage}[l]{0.48\linewidth}
\epsfig{file=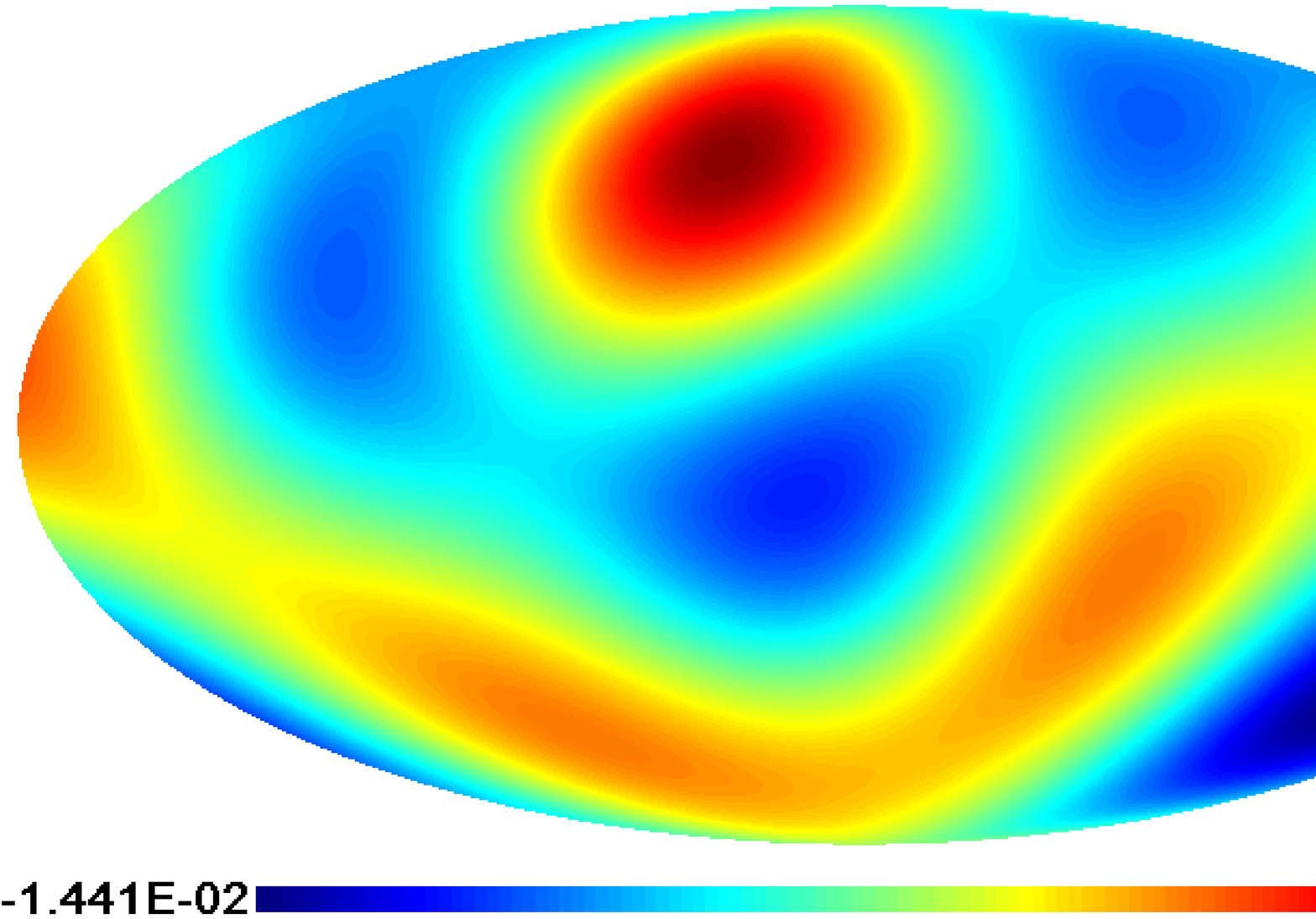,height=0.2\textheight,width=\linewidth}
\end{minipage}
\begin{minipage}[l]{0.48\linewidth}
\epsfig{file=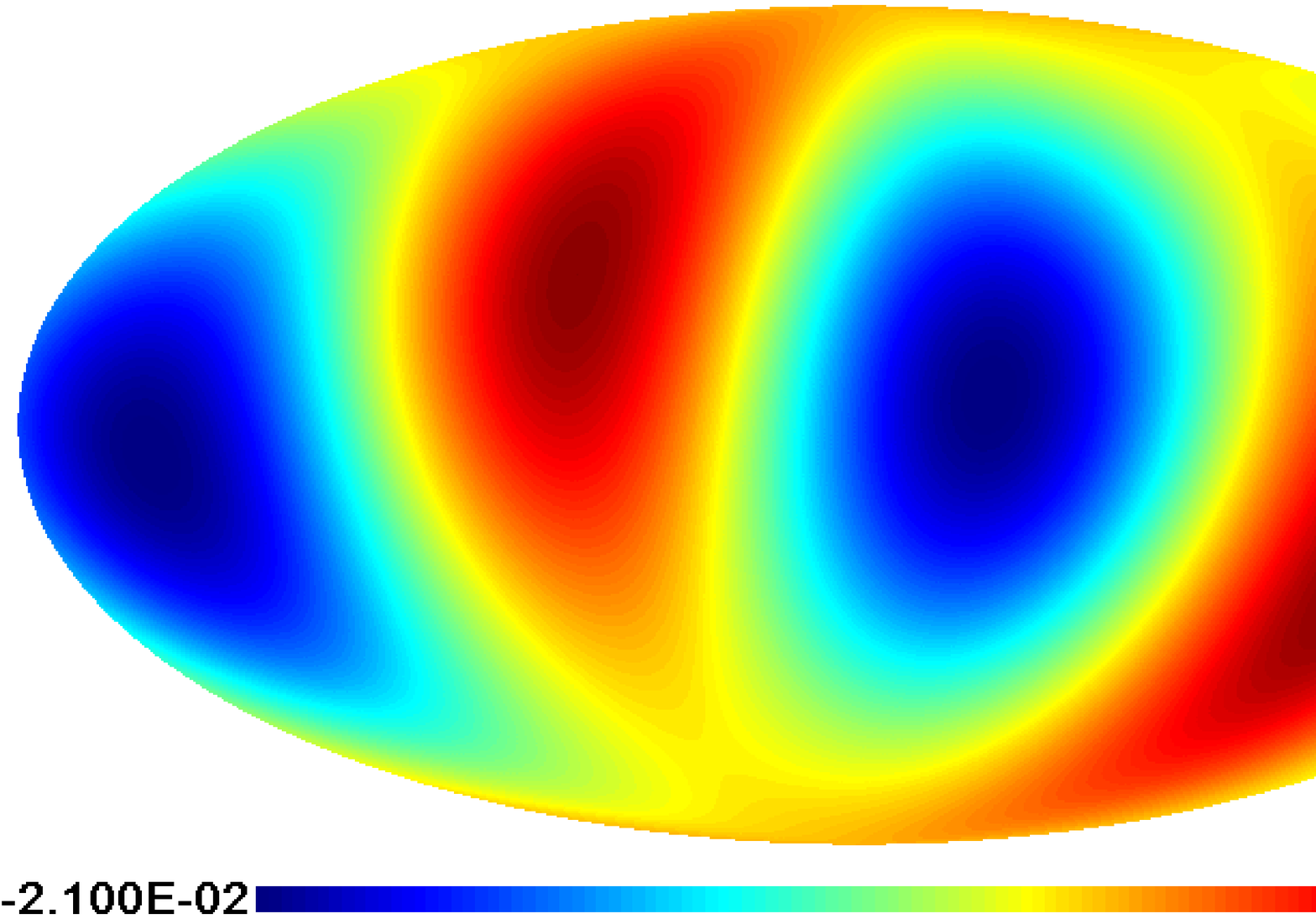,height=0.2\textheight,width=\linewidth}
\end{minipage}
\begin{minipage}[l]{0.48\linewidth}
\epsfig{file=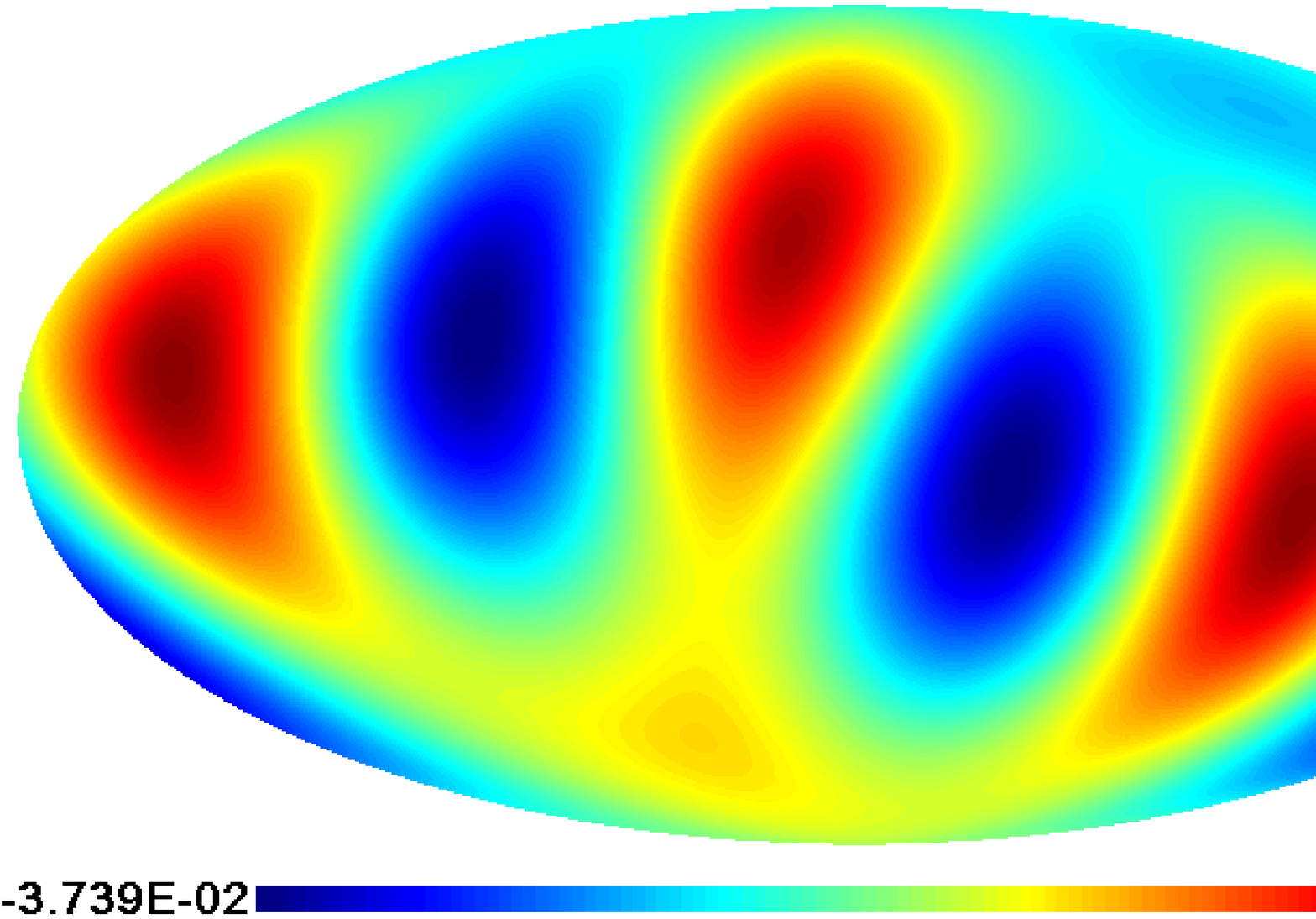,height=0.2\textheight,width=\linewidth}
\end{minipage}
\begin{minipage}[l]{0.48\linewidth}
\epsfig{file=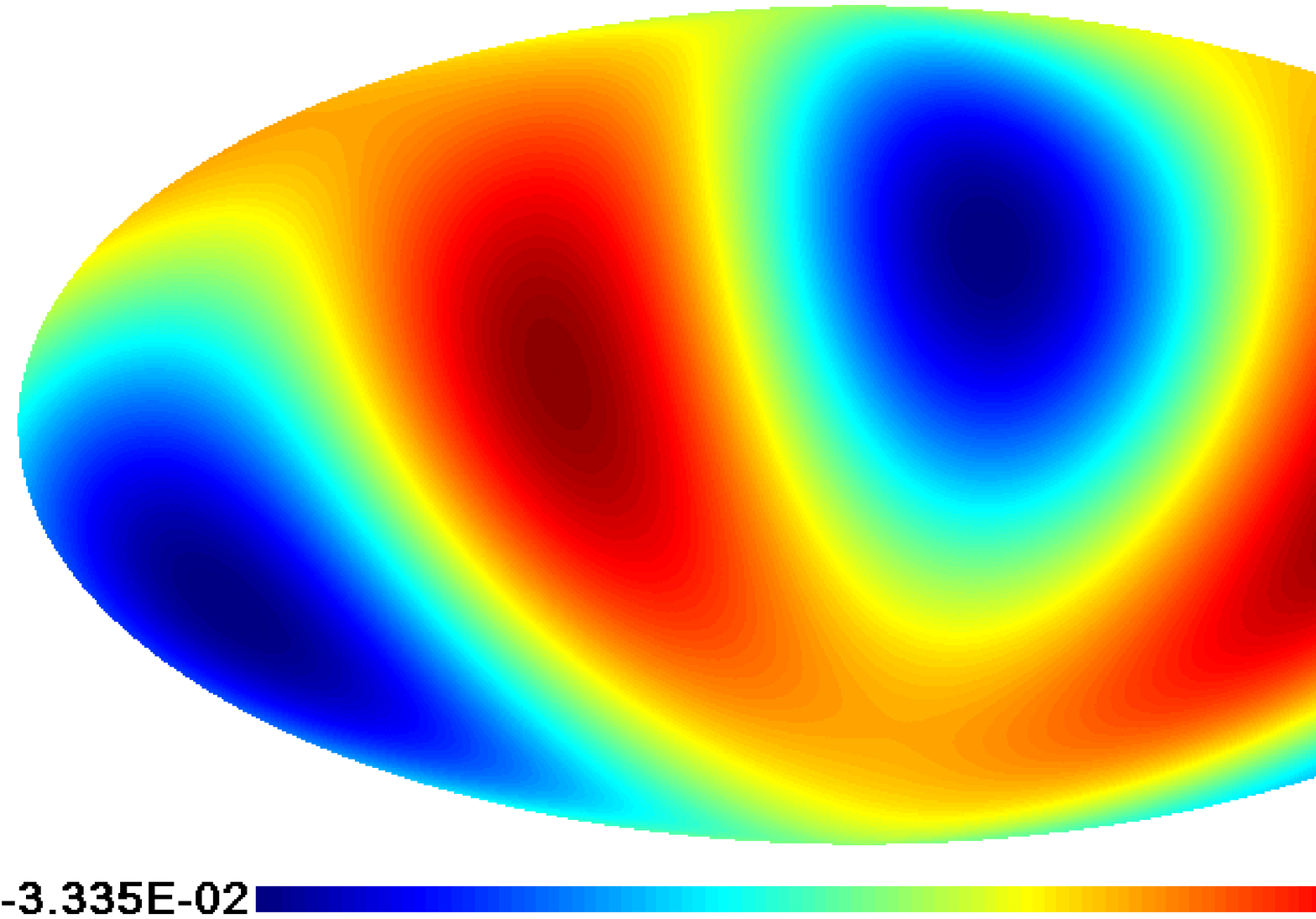,height=0.2\textheight,width=\linewidth}
\end{minipage}
\begin{minipage}[l]{0.48\linewidth}
\epsfig{file=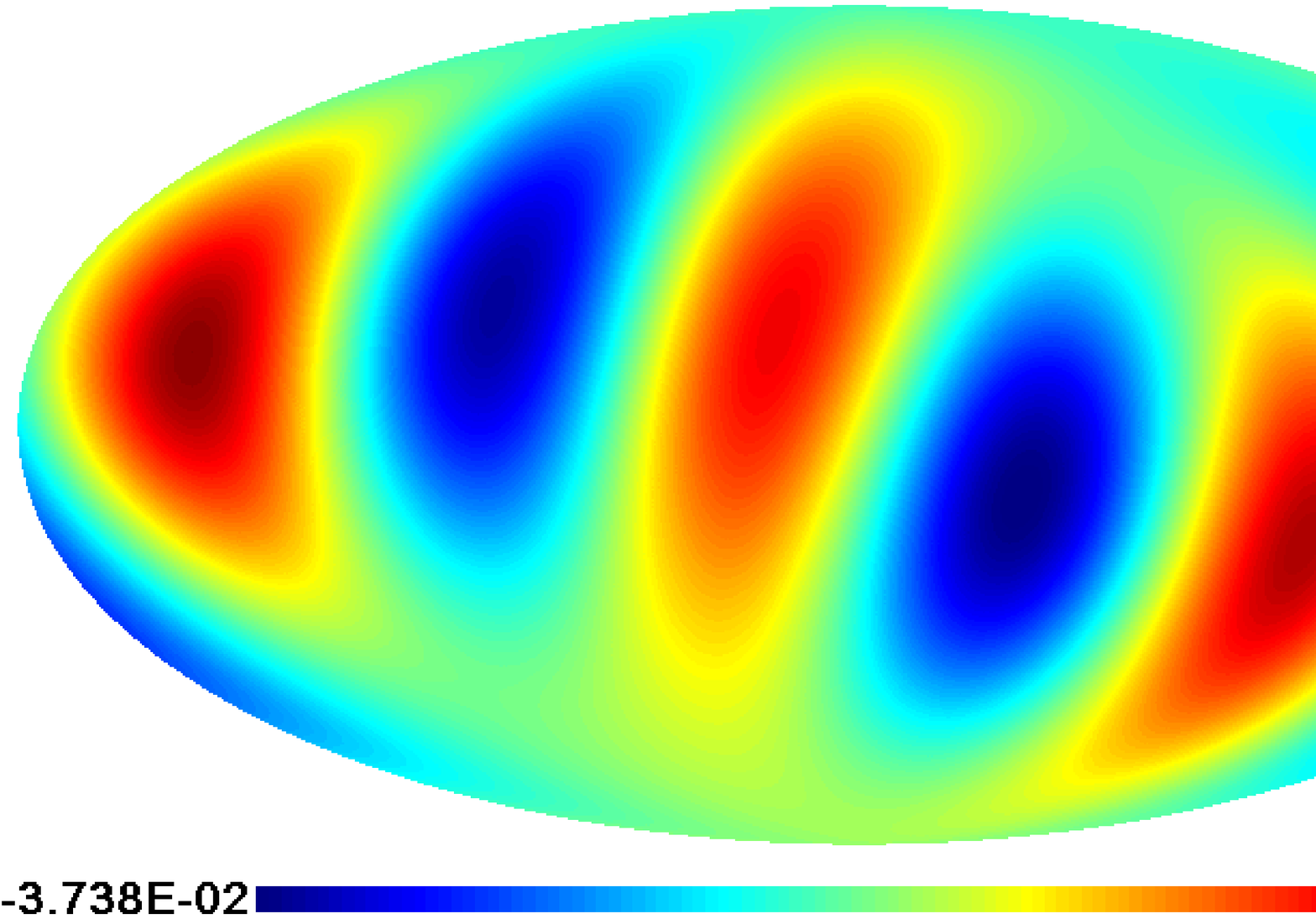,height=0.2\textheight,width=\linewidth}
\end{minipage}
\caption{(Top) The quadrupole (left) and octopole (right) in mK of the
  estimated local ISW effect. (Centre) The quadrupole (left) and
  octopole (right) in mK of the WMAP ILC map. (Bottom) The
  quadrupole (left) and octopole (right) in mK of the ILC map after
  subtracting the local ISW contribution. \label{fig:low_l_maps}}
\end{figure*}

We now consider the effect of removing the local ISW signal on each of
the large scale CMB anomalies. We use the year-3 WMAP ILC map
(hereafter ILC) and the CMB map computed by \citet{deOliveira_2006}
from the year-3 WMAP data (hereafter OT). 
We have removed the kinetic Doppler
dipole and quadrupole contributions due
to the motion of the Sun with respect to the CMB frame.

\subsection{Low quadrupole power}

Figure \ref{fig:low_l_maps} shows the $\ell=2$ and $\ell=3$ multipoles
of the estimated local ISW signal, the WMAP ILC map and the WMAP ILC
map after removal of the estimated ISW signal.  We immediately note a
significant increase in the temperature range of the quadrupole map
after ISW removal and an alteration in the positions of the
temperature extrema for this multipole; in contrast, the octopole is
largely unchanged.  We compute the power of the quadrupole for these
maps, together with the foreground-cleaned map of
\citet{deOliveira_2006} before and after ISW subtraction. The results
and their significance given the WMAP best-fitting model prediction
are given in Table \ref{table:quad_power}. We also repeat the exercise
for the CMB maps with the rescaled version of the local ISW prediction
removed.  The results in Table \ref{table:quad_power} show that the
low value of the quadrupole power ceases to be significant when we
consider the maps with either the original or the rescaled versions of
the estimated local ISW signal removed. This is one of the main results
of this paper, and its broader significance is discussed in the 
concluding section.

\begin{table}
\centering
\begin{tabular}{c|c|c} \hline
Map  & Quadrupole Power / $\mu$K$^2$ & Probability  \\ \hline
ILC & $250.6$ & $3.5\%$    \\
OT & $214.3$ & $2.4\%$   \\
ILC -- ISW  & $600.7$ & $19.8\%$   \\
OT -- ISW  & $608.2$ & $20.1\%$   \\
%ILC -- ISW (rescaled) & $372.7$ & $8.0\%$   \\
%OT -- ISW (rescaled) & $359.3$ & $7.5\%$   \\ \hline
% Using cosmological model from letter to calculate rescaling factors
ILC -- ISW (rescaled) & $372.2$ & $8.0\%$   \\
OT -- ISW (rescaled) & $358.8$ & $7.5\%$   \\ \hline
\end{tabular}
\caption{The quadrupole power for the WMAP ILC and OT maps before and
  after subtraction of the estimated local ISW signal for $z<0.3$,
  together with its significance given the best-fitting year-3 WMAP
  prediction. This is a one-tailed probability of having a quadrupole
  as low or lower than observed, given the distribution
  of quadrupole power ($\chi^2$ with 5 d.f.). 
  We note that the quadrupole power is no longer
  unusually low after subtraction of the ISW signal. \label{table:quad_power}}
\end{table}

\subsection{Quadrupole/octopole alignment}

In order to quantify the alignment between the quadrupole and octopole, we have used the maximum angular dispersion technique of
\citet{deOliveira} to identify preferred axes for the quadrupole, $\mathbf{\hat{n}}_{2}$, and
octopole, $\mathbf{\hat{n}}_{3}$. Thus we search over directions $\mathbf{\hat{n}}$ to maximise
the quantity $\sum_{m}{m^{2}|a_{\ell m}|^{2}}$ in a coordinate system
with $\mathbf{\hat{n}}$ as the direction of the $z$-axis. Our results
are given in Table \ref{table:quad_oct_align}. We see that there is no significant alignment remaining in
either the ILC or OT map after removal of either the estimated local ISW
signal or the rescaled version.

\begin{table}
\centering
\begin{tabular}{c|c|c|c} \hline
Map  & $\mathbf{\hat{n}}_{2}\cdot\mathbf{\hat{n}}_{3}$ & Separation & Probability  \\ \hline
ILC & $0.9991$ & $2.4^{\circ}$  & $0.09\%$   \\
OT & $0.9881$ & $8.9^{\circ}$ & $1.2\%$  \\
ILC -- ISW & $0.7548$ & $41.0^{\circ}$ & $24.5\%$   \\
OT -- ISW & $0.6712$ & $47.8^{\circ}$ & $32.9\%$   \\
%ILC -- ISW (rescaled) & $0.8873$ & $27.5^{\circ}$ & $11.3\%$   \\ 
%OT -- ISW (rescaled) & $0.8158$ & $35.3^{\circ}$ & $19.4\%$   \\ \hline
% Using cosmological model from letter to calculate rescaling factors
ILC -- ISW (rescaled) & $0.8888$ & $27.3^{\circ}$ & $11.1\%$   \\ 
OT -- ISW (rescaled) & $0.8183$ & $35.1^{\circ}$ & $19.2\%$   \\ \hline
\end{tabular}
\caption{The scalar product of the preferred axes of the quadrupole
  and octopole together with their angular separation and the one-tailed
  probability of such a good separation occurring by random alignment. Results are shown for the
  ILC and OT maps before and after removal of both the estimated local ISW
  signal and the rescaled version. \label{table:quad_oct_align}}
\end{table}

\begin{figure*}
\begin{minipage}[l]{0.48\linewidth}
\epsfig{file=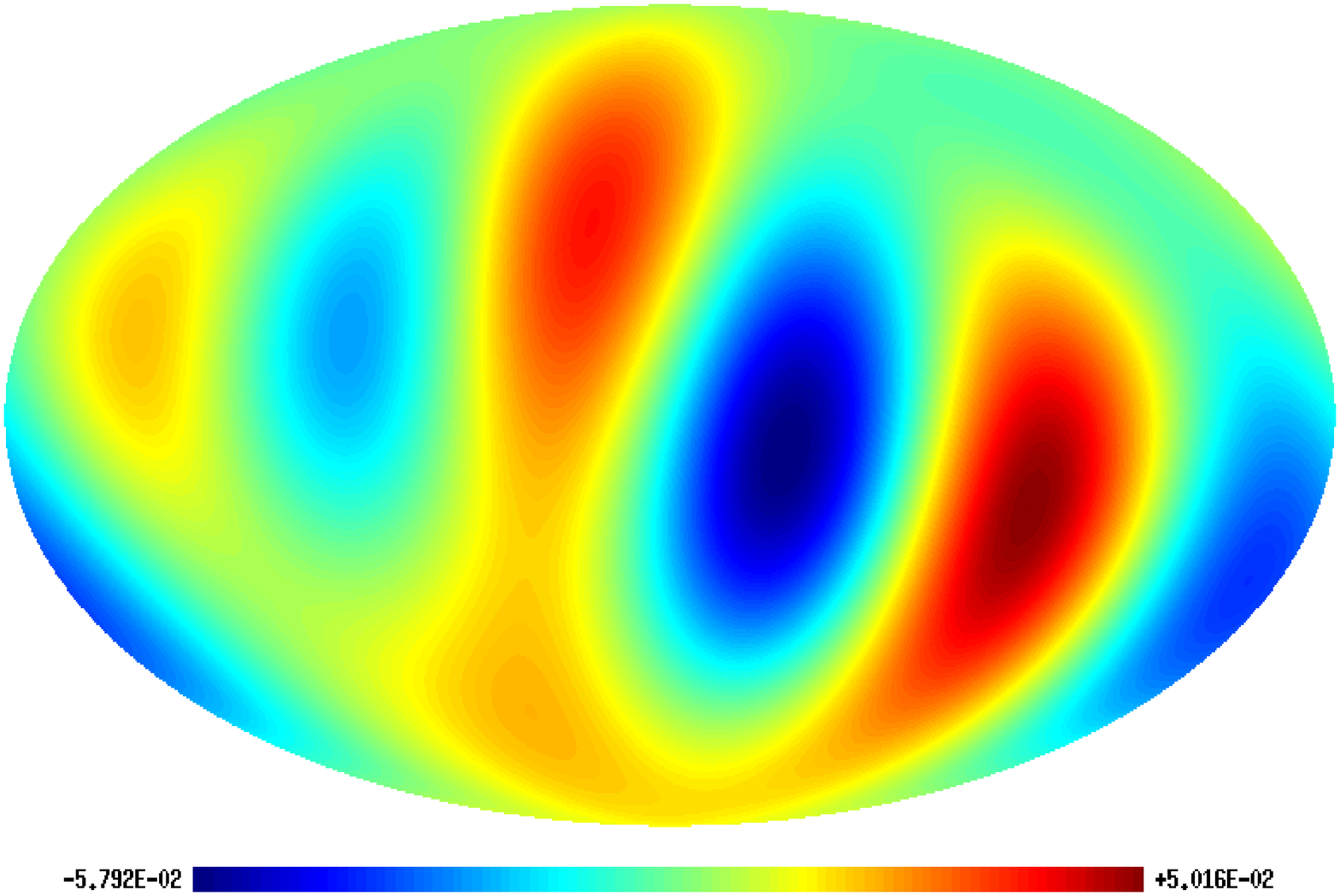,height=0.2\textheight,width=\linewidth}
\end{minipage}
\begin{minipage}[r]{0.48\linewidth}
\epsfig{file=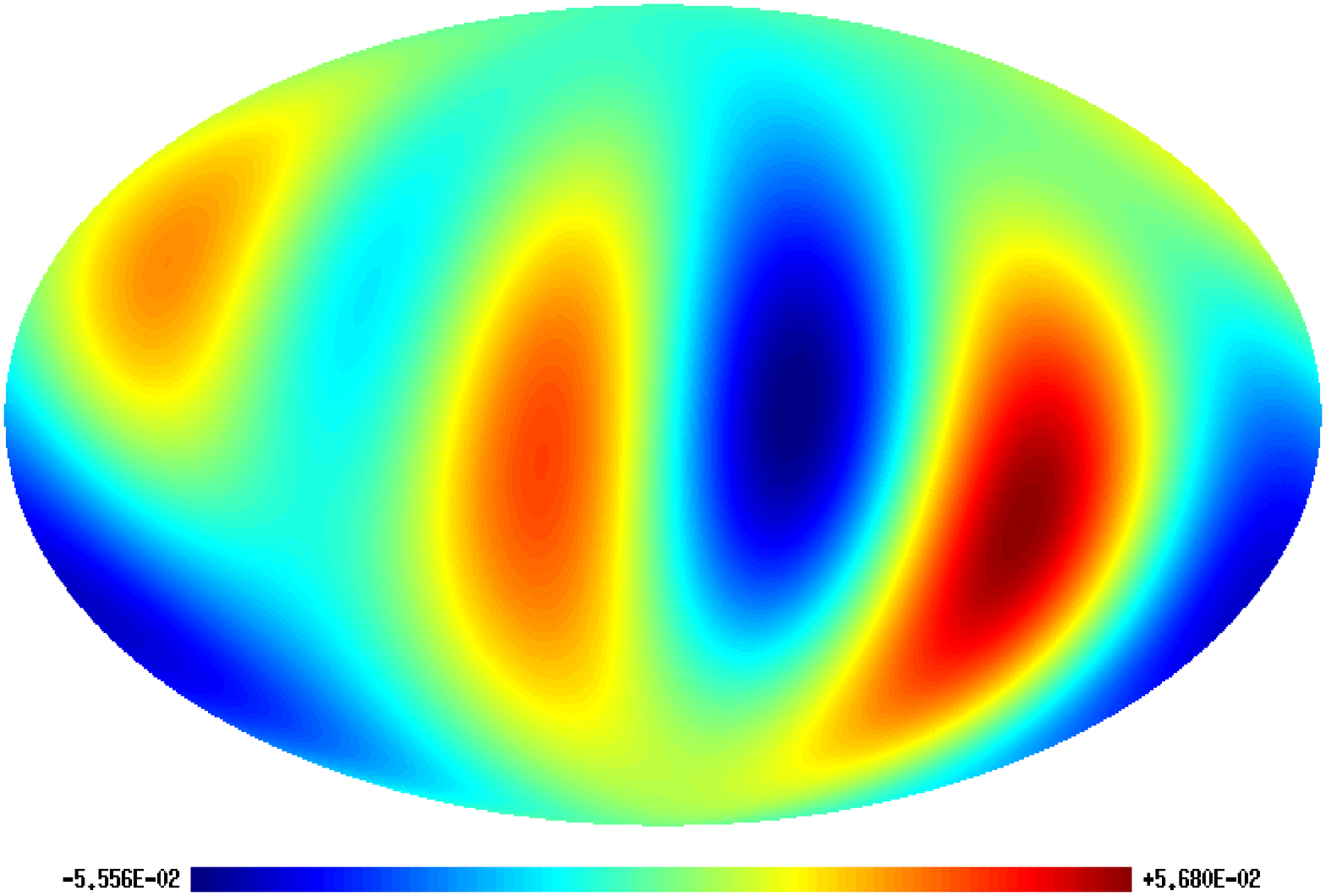,height=0.2\textheight,width=\linewidth}
\end{minipage}
\caption{The $\ell=2+3$ multipoles of the year-3 ILC map (left) and
  the ILC map after subtraction of the estimated local ISW signal
  (right) in mK; the black line shows the ecliptic plane. We note that
  the power asymmetry persists, with the largest temperature extrema
  seen south of the ecliptic plane, but that the ecliptic no longer
  lies along a node line of the temperature
  distribution.\label{fig:north_south}}
\end{figure*}

\begin{table}
\centering
\begin{tabular}{c|c|c} \hline
Map  & `$t$' value & Probability  \\ \hline
ILC & $0.9213$ & ~$17\%$    \\
OT & $0.9269$ & ~$15\%$  \\
ILC -- ISW  & $0.9841$ & $1.6\%$   \\
OT -- ISW  & $0.9653$ & $5.0\%$   \\
%ILC -- ISW (rescaled) & $0.9675$ & $4.9\%$  \\
%OT -- ISW (rescaled) & $0.9530$ & $8.0\%$  \\ \hline
% Using cosmological model from letter to calculate rescaling factors
ILC -- ISW (rescaled) & $0.9661$ & $4.9\%$  \\
OT -- ISW (rescaled) & $0.9524$ & $8.0\%$  \\ \hline
\end{tabular}
\caption{The $t$ value for the octopoles of each of our usual maps
  together with the one-tailed probability of such an extreme value occurring by chance, as
  determined from 10000 Monte-Carlo simulations of $a_{3m}$
  coefficients. Removal of the estimated local ISW effect actually
  makes the octopole more planar.\label{table:planar_oct}}
\end{table}

\subsection{Planarity of the octopole}

Fig. \ref{fig:low_l_maps} shows that the octopole is not particularly
affected by the subtraction of the estimated local ISW
effect. Evaluating the `$t$' statistic of \citet{deOliveira} 
\[
t = \max_{\mathbf{\hat{n}}}\frac{|a_{3 -3}|^{2} +
  |a_{33}|^{2}}{\sum_{m}{|a_{\ell m}|^{2}}}
\]
as a measure of planarity, we find the results given in
Table \ref{table:planar_oct}.
The significance of the `$t$' statistic is evaluated by
considering 10000 simulated sets of $a_{\ell m}$ data and determining
the frequency of such `$t$' values occurring by chance (see de
Oliveira-Costa et al. 2004 for further details). The ISW subtracted
maps have octopoles that are slightly more planar than the
corresponding ILC or OT maps, i.e. removal of the estimated local ISW
effect seems to make the octopole more planar. The probability of such
planarity occurring by chance varies depending on the CMB map which is
used, with probabilities of $\sim 2-5\%$ for the full local ISW
subtraction and $\sim 5-8\%$ for the rescaled ISW subtraction.

\subsection{North-South asymmetry}

Fig. \ref{fig:north_south} shows the sum of the quadrupole and
octopole for the ILC map, before and after ISW subtraction. An
asymmetry in power between the hemispheres north and south of the
ecliptic remains after removal of the estimated ISW signal, but the
ecliptic no longer lies along a null of the temperature
distribution. The significance of the power asymmetry and the
coincidence of the ecliptic with a node line across 1/3 of the sky was
found by \citet{Schwarz_2004} to be unusual at the $95\%$ level. We
expect that the breaking of this coincidence with the ecliptic that
occurs after ISW removal would reduce the significance of the anomaly
much further.

\subsection{Cold spot}

The `cold spot' is centred at $\ell=207.8\deg, b=-56.3\deg$ and has
scale $\sim 10\deg$. This feature was identified by
\citet{Vielva_coldspot} using Spherical Mexican Hat Wavelets in
the year-1 WMAP CMB data and confirmed in the year-3 WMAP data
\citep{Cruz_coldspot_2007}. \citet{Rudnick_NVSS_coldspot}
identified the region of the cold spot with an underdensity of NVSS
radio sources and therefore suggested that the cold spot is due to the
ISW contribution of a large void at redshift $z\sim1$. However,
\citet{Smith_Huterer} find no evidence for such an underdensity of
NVSS sources. We have evaluated the CMB temperature in the
vicinity of the cold spot 
before and after the subtraction of the estimated local
ISW signal; the pixels in the vicinity of the cold spot
become on average $7\mu$K warmer after ISW removal.
This changes the formal significance level of the feature
from 1.85\% to 2.5\%.

If the cold spot were due to the ISW contribution of a
large void at $z\sim1$, removal of the estimated local ISW signal
slightly reduces the size of the necessary void to around $95\%$ of
the current estimate of about $105\mpcoh$ at $z\simeq 1$
\citep{Rudnick_NVSS_coldspot}.  A void of this size 
would remain anomalously large.

\section{Conclusions}

\label{sec:concl}

We have estimated the local ISW signal to $z=0.3$ based on large scale
structure data from 2MASS and SuperCOSMOS. On subtracting this predicted foreground
from CMB data, we find that many large-scale CMB anomalies are
alleviated. In particular the quadrupole power is raised and the
alignment between the $\ell=2$ and $\ell=3$ multipoles is broken. We
have checked that these results are not sensitive to the precise
amplitude of the low multipoles of the estimated ISW signal
by considering in addition a rescaled version of the ISW map, chosen
to have the same phases as the original estimation but a power
spectrum equal to that expected for the local ISW effect with
$\Omega_m=0.3$. On subtraction of the estimated ISW foreground, we
also find that the co-incidental alignment of the ecliptic with a node
line of the $\ell=2+3$ multipoles of the CMB is broken (although the
north-south asymmetry in the signal remains) and that the temperature
in the vicinity of the non-Gaussian cold spot is slightly raised
(although the change in significance is not substantial).
Of the large-scale anomalies considered, the planarity
of the octopole remains most unusual, and in fact becomes more
significant on removal of the ISW signal -- but $\sim5\%$ significance
is hardly compelling evidence for any anomaly.

These results thus demonstrate that removal of the local ISW effect can
mitigate the anomalous signals that have been claimed on large scales
in the CMB. Our estimate of the CMB signal that originates at
$z>0.3$ has a character that is a better match to the expected
Gaussian random field. 
The harmonic-space anomalies investigated here are closely
related to the issue of lack of large-scale angular correlations,
and one might expect that ISW subtraction would also remove this
problem; \citet{Efstathiou_2009} have 
used the data from the present paper to show that this is indeed the case.

It should be borne in mind that we have only considered
the ISW signal for $0<z<0.3$ in this work, which we estimate to
contain around $40\%$ of the total rms ISW signal at low multipoles,
and therefore there will be further substantial influence on the
observed CMB from the ISW effect at higher redshifts:
an upper limit of $z<1.3$ would yield $>90\%$ of the signal.
Future imaging surveys will provide photometric redshifts to this
depth (approximately an AB magnitude limit of $r=24$), but extensive
sky coverage will be hard to obtain; the Dark Energy Survey
\citep{DES_whitepaper} will cover 5000 deg$^2$ by about 2015, but
the full extragalactic sky will probably have to await results from LSST
\citep{LSST_summary} in perhaps 2020,
or ultimately ESA's Euclid mission \citep{Euclid_summary},
which is currently scheduled to finish in about 2023.

Our work nevertheless
demonstrates that it is important to estimate the contribution of ISW
secondary anisotropies to the CMB before concluding that the
large scale intrinsic CMB signal is anomalous or otherwise. 
In this respect, we agree with previous workers who have argued
that foregrounds from large-scale structure might affect the CMB
anomalies: \citet{Rakic_Rasanen} looked at the nonlinear Rees-Sciama
effect and \citet{Abramo_2006} the Sunyaev-Zeldovich effect. However, it is
well established that the ISW signal is much the largest of these
foregrounds on the relevant scales \citep[e.g.][]{Cooray_and_Sheth2002}, and so ISW
removal should dominate any change in the character of the CMB sky.

Of course, it can be argued that we have replaced one puzzle by another:
the intrinsic CMB should be statistically isotropic and Gaussian, and the large-angle
ISW signal should very nearly have the same character. At the lowest
multipoles, the prediction of the standard cosmology is that the power
from each of these components should be very nearly equal
\citep[see Figure 62 from][]{Cooray_and_Sheth2002}.
How could the sum of two uncorrelated Gaussian fields {\it reduce\/} the
total quadrupole power, and introduce non-Gaussian correlations
at higher multipoles? Within the standard model, such an
outcome would have to be a fluke, and it is true that
some authors \citep[e.g.][]{Efstathiou_2004} claim
that the statistical significance
of the claimed anomalies is not overwhelming.

If we reject the idea of a fluke, then either the intrinsic
CMB or the foreground would have to be anomalous, unless
a new mechanism can be found to correlate these two components.
Regarding non-standard foregrounds,
\citet{Gordon_2005} showed that it is hard to produce the
complete set of claimed anomalies by adding uncorrelated signals
even if the foreground component is made explicitly anisotropic;
rather, they claim that one would need a multiplicative anisotropy in which
the temperature variance has a large-scale variation over the sky.
At higher multipoles than considered here, there is accumulating
evidence that just such an effect exists, with both a dipole
and quadrupole modulation \citep[e.g.][]{Hanson_Lewis_ecliptic,Hoftuft09}.
If such effects are not low-level instrumental systematics, then a
radical new ingredient in cosmological structure
formation is required; but in either case, there is no evidence that these
modulations have any relation to the claimed large-angle anomalies.
Our position on these is that 
exotic mechanisms for explaining CMB anomalies as a real 
physical effect \citep[e.g.][]{Luminet_2003}, 
tend to apply only to the intrinsic anisotropies that originate at
$z\simeq 1100$, so that we should try to isolate these 
from foreground emission as cleanly as possible. We have tried to
perform this task in the present paper by removing part of the
ISW foreground, with the result that the estimate of the
intrinsic CMB no longer appears unusual in its
large-scale properties. This reduces the motivation for revisions to
the standard theory for the origin of CMB fluctuations, even if it
leaves the door open for a more elaborate model that could
introduce a correlation between the intrinsic CMB and
foreground anisotropies.

%%%%%%%%%%%%%%%%%%%%%%%%%%%%%%%%%%%%%%%%%%%%%%%%%%%%%%%%%%%%%%%%%%%%%%%%%%%%%%%%%

\section*{Acknowledgements}

CLF was supported by a PPARC PhD studentship. We thank Kate
Land for much helpful correspondence. This research has made use of
optical data obtained from the SuperCOSMOS Science Archive, prepared
and hosted by the IfA's Wide Field Astronomy Unit, consisting of
scanned survey plates from the UK Schmidt Telescope and The Palomar
Observatory Sky Survey (POSS-II). This publication also makes use of
data products from the Two Micron All Sky Survey, which is a joint
project of the University of Massachusetts and the Infrared Processing
and Analysis Center/California Institute of Technology, funded by the
National Aeronautics and Space Administration and the National Science
Foundation. \phantom{\citet{Gorski_HEALPix}}

\setlength{\bibhang}{2.0em}
\setlength\labelwidth{0.0em}
\bibliography{biblio}

\begin{thebibliography}{}

\bibitem[\protect\citeauthoryear{{Abramo}, {Sodr{\'e}} \& {Wuensche}}{{Abramo}
  et~al.}{2006}]{Abramo_2006}
{Abramo} L.~R.,  {Sodr{\'e}} L.~J.,    {Wuensche} C.~A.,  2006, \prd, 74,
  083515

\bibitem[\protect\citeauthoryear{{Afshordi}, {Loh} \& {Strauss}}{{Afshordi}
  et~al.}{2004}]{Afshordi}
{Afshordi} N.,  {Loh} Y.-S.,    {Strauss} M.~A.,  2004, \prd, 69, 083524

\bibitem[\protect\citeauthoryear{{Cooray} \& {Sheth}}{{Cooray} \&
  {Sheth}}{2002}]{Cooray_and_Sheth2002}
{Cooray} A.,  {Sheth} R.,  2002, Physics Reports, 372, 1

\bibitem[\protect\citeauthoryear{{Copi}, {Huterer}, {Schwarz} \&
  {Starkman}}{{Copi} et~al.}{2006}]{Copi_quad+oct}
{Copi} C.~J.,  {Huterer} D.,  {Schwarz} D.~J.,    {Starkman} G.~D.,  2006,
  \mnras, 367, 79

\bibitem[\protect\citeauthoryear{{Copi}, {Huterer}, {Schwarz} \&
  {Starkman}}{{Copi} et~al.}{2009}]{Copi_2009}
{Copi} C.~J.,  {Huterer} D.,  {Schwarz} D.~J.,    {Starkman} G.~D.,  2009,
  \mnras, 399, 295

\bibitem[\protect\citeauthoryear{{Crittenden} \& {Turok}}{{Crittenden} \&
  {Turok}}{1996}]{Crittenden_Turok_1996}
{Crittenden} R.~G.,  {Turok} N.,  1996, Physical Review Letters, 76, 575

\bibitem[\protect\citeauthoryear{{Cruz}, {Cay{\'o}n},
  {Mart{\'{\i}}nez-Gonz{\'a}lez}, {Vielva} \& {Jin}}{{Cruz}
  et~al.}{2007}]{Cruz_coldspot_2007}
{Cruz} M.,  {Cay{\'o}n} L.,  {Mart{\'{\i}}nez-Gonz{\'a}lez} E.,  {Vielva} P.,
   {Jin} J.,  2007, \apj, 655, 11

\bibitem[\protect\citeauthoryear{{de Oliveira-Costa} \& {Tegmark}}{{de
  Oliveira-Costa} \& {Tegmark}}{2006}]{deOliveira_2006}
{de Oliveira-Costa} A.,  {Tegmark} M.,  2006, \prd, 74, 023005

\bibitem[\protect\citeauthoryear{{de Oliveira-Costa}, {Tegmark}, {Zaldarriaga}
  \& {Hamilton}}{{de Oliveira-Costa} et~al.}{2004}]{deOliveira}
{de Oliveira-Costa} A.,  {Tegmark} M.,  {Zaldarriaga} M.,    {Hamilton} A.,
  2004, \prd, 69, 063516

\bibitem[\protect\citeauthoryear{{DES Collaboration}}{{DES
  Collaboration}}{2005}]{DES_whitepaper}
{DES Collaboration} 2005, arXiv:astro-ph/0510346

\bibitem[\protect\citeauthoryear{{Efstathiou}}{{Efstathiou}}{2004}]{Efstathiou%
_2004}
{Efstathiou} G.,  2004, \mnras, 348, 885

\bibitem[\protect\citeauthoryear{{Efstathiou}, {Ma} \& {Hanson}}{{Efstathiou}
  et~al.}{2009}]{Efstathiou_2009}
{Efstathiou} G.,  {Ma} Y.,    {Hanson} D.,  2009, arXiv:0911.5399

\bibitem[\protect\citeauthoryear{{Eriksen}, {Hansen}, {Banday}, {G{\'o}rski} \&
  {Lilje}}{{Eriksen} et~al.}{2004}]{Eriksen_northsouth}
{Eriksen} H.~K.,  {Hansen} F.~K.,  {Banday} A.~J.,  {G{\'o}rski} K.~M.,
  {Lilje} P.~B.,  2004, \apj, 605, 14

\bibitem[\protect\citeauthoryear{{Francis}}{{Francis}}{2008}]{Francis_PHD}
{Francis} C.~L.,  2008, PhD thesis, University of Edinburgh

\bibitem[\protect\citeauthoryear{{Francis} \& {Peacock}}{{Francis} \&
  {Peacock}}{2009}]{Francis_ISW}
{Francis} C.~L.,  {Peacock} J.~A.,  2009, arXiv:0909.2494

\bibitem[\protect\citeauthoryear{{Gordon}, {Hu}, {Huterer} \&
  {Crawford}}{{Gordon} et~al.}{2005}]{Gordon_2005}
{Gordon} C.,  {Hu} W.,  {Huterer} D.,    {Crawford} T.,  2005, \prd, 72, 103002

\bibitem[\protect\citeauthoryear{{G{\'o}rski}, {Hivon}, {Banday}, {Wandelt},
  {Hansen}, {Reinecke} \& {Bartelmann}}{{G{\'o}rski}
  et~al.}{2005}]{Gorski_HEALPix}
{G{\'o}rski} K.~M.,  {Hivon} E.,  {Banday} A.~J.,  {Wandelt} B.~D.,  {Hansen}
  F.~K.,  {Reinecke} M.,    {Bartelmann} M.,  2005, \apj, 622, 759

\bibitem[\protect\citeauthoryear{{Hambly}, {Davenhall}, {Irwin} \&
  {MacGillivray}}{{Hambly} et~al.}{2001}]{Hambly_SuperCOSMOS}
{Hambly} N.~C.,  {Davenhall} A.~C.,  {Irwin} M.~J.,    {MacGillivray} H.~T.,
  2001, \mnras, 326, 1315

\bibitem[\protect\citeauthoryear{{Hanson} \& {Lewis}}{{Hanson} \&
  {Lewis}}{2009}]{Hanson_Lewis_ecliptic}
{Hanson} D.,  {Lewis} A.,  2009, \prd, 80, 063004

\bibitem[\protect\citeauthoryear{{Hinshaw}, {Banday}, {Bennett}, {Gorski},
  {Kogut}, {Lineweaver}, {Smoot} \& {Wright}}{{Hinshaw}
  et~al.}{1996}]{Hinshaw_1996}
{Hinshaw} G.,  {Banday} A.~J.,  {Bennett} C.~L.,  {Gorski} K.~M.,  {Kogut} A.,
  {Lineweaver} C.~H.,  {Smoot} G.~F.,    {Wright} E.~L.,  1996, \apjl, 464, L25

\bibitem[\protect\citeauthoryear{{Hoftuft}, {Eriksen}, {Banday}, {G{\'o}rski},
  {Hansen} \& {Lilje}}{{Hoftuft} et~al.}{2009}]{Hoftuft09}
{Hoftuft} J.,  {Eriksen} H.~K.,  {Banday} A.~J.,  {G{\'o}rski} K.~M.,  {Hansen}
  F.~K.,    {Lilje} P.~B.,  2009, \apj, 699, 985

\bibitem[\protect\citeauthoryear{{Ivezic}, {Tyson}, {Allsman}, {Andrew},
  {Angel} \& {for the LSST Collaboration}}{{Ivezic}
  et~al.}{2008}]{LSST_summary}
{Ivezic} Z.,  {Tyson} J.~A.,  {Allsman} R.,  {Andrew} J.,  {Angel} R.,    {for
  the LSST Collaboration} 2008, arXiv:0805.2366

\bibitem[\protect\citeauthoryear{{Jarrett}}{{Jarrett}}{2004}]{Jarrett_2MASS}
{Jarrett} T.,  2004, Publications of the Astronomical Society of Australia, 21,
  396

\bibitem[\protect\citeauthoryear{{Komatsu} et~al.,}{{Komatsu}
  et~al.}{2009}]{KomatsuWMAP09}
{Komatsu} E.,  et~al., 2009, \apjs, 180, 330

\bibitem[\protect\citeauthoryear{{Land} \& {Magueijo}}{{Land} \&
  {Magueijo}}{2005}]{Land_Magueijo_2005}
{Land} K.,  {Magueijo} J.,  2005, Physical Review Letters, 95, 071301

\bibitem[\protect\citeauthoryear{{Laureijs}}{{Laureijs}}{2009}]{Euclid_summary}
{Laureijs} R.,  2009, arXiv:0912.0914

\bibitem[\protect\citeauthoryear{{Luminet}, {Weeks}, {Riazuelo}, {Lehoucq} \&
  {Uzan}}{{Luminet} et~al.}{2003}]{Luminet_2003}
{Luminet} J.-P.,  {Weeks} J.~R.,  {Riazuelo} A.,  {Lehoucq} R.,    {Uzan}
  J.-P.,  2003, \nat, 425, 593

\bibitem[\protect\citeauthoryear{{Martinez-Gonzalez}, {Sanz} \&
  {Silk}}{{Martinez-Gonzalez} et~al.}{1990}]{Martinez-Gonzalez_Sanz_Silk}
{Martinez-Gonzalez} E.,  {Sanz} J.~L.,    {Silk} J.,  1990, \apjl, 355, L5

\bibitem[\protect\citeauthoryear{{Peacock} et~al.,}{{Peacock}
  et~al.}{2010}]{Peacock_photoz}
{Peacock} J.~A.,  et~al., 2010, in prep.

\bibitem[\protect\citeauthoryear{Rakic, Rasanen \& Schwarz}{Rakic
  et~al.}{2006}]{Rakic_Rasanen}
Rakic A.,  Rasanen S.,    Schwarz D.~J.,  2006, \mnras, 369, L27

\bibitem[\protect\citeauthoryear{{Rudnick}, {Brown} \& {Williams}}{{Rudnick}
  et~al.}{2007}]{Rudnick_NVSS_coldspot}
{Rudnick} L.,  {Brown} S.,    {Williams} L.~R.,  2007, \apj, 671, 40

\bibitem[\protect\citeauthoryear{{Schlegel}, {Finkbeiner} \&
  {Davis}}{{Schlegel} et~al.}{1998}]{Schlegel_dust}
{Schlegel} D.~J.,  {Finkbeiner} D.~P.,    {Davis} M.,  1998, \apj, 500, 525

\bibitem[\protect\citeauthoryear{{Schwarz}, {Starkman}, {Huterer} \&
  {Copi}}{{Schwarz} et~al.}{2004}]{Schwarz_2004}
{Schwarz} D.~J.,  {Starkman} G.~D.,  {Huterer} D.,    {Copi} C.~J.,  2004,
  Physical Review Letters, 93, 221301

\bibitem[\protect\citeauthoryear{{Smith} \& {Huterer}}{{Smith} \&
  {Huterer}}{2010}]{Smith_Huterer}
{Smith} K.~M.,  {Huterer} D.,  2010, \mnras, 403, 2

\bibitem[\protect\citeauthoryear{{Spergel} et~al.,}{{Spergel}
  et~al.}{2003}]{Spergel_lowl2}
{Spergel} D.~N.,  et~al., 2003, \apjs, 148, 175

\bibitem[\protect\citeauthoryear{{Tegmark}, {de Oliveira-Costa} \&
  {Hamilton}}{{Tegmark} et~al.}{2003}]{Tegmark_l2l3align}
{Tegmark} M.,  {de Oliveira-Costa} A.,    {Hamilton} A.~J.,  2003, \prd, 68,
  123523

\bibitem[\protect\citeauthoryear{{Vielva}, {Mart{\'{\i}}nez-Gonz{\'a}lez},
  {Barreiro}, {Sanz} \& {Cay{\'o}n}}{{Vielva} et~al.}{2004}]{Vielva_coldspot}
{Vielva} P.,  {Mart{\'{\i}}nez-Gonz{\'a}lez} E.,  {Barreiro} R.~B.,  {Sanz}
  J.~L.,    {Cay{\'o}n} L.,  2004, \apj, 609, 22

\bibitem[\protect\citeauthoryear{{Zaroubi}, {Hoffman}, {Fisher} \&
  {Lahav}}{{Zaroubi} et~al.}{1995}]{Zaroubi_1995}
{Zaroubi} S.,  {Hoffman} Y.,  {Fisher} K.~B.,    {Lahav} O.,  1995, \apj, 449,
  446

\end{thebibliography}

%\appendix

%\section{}

%\subsection{}

\bsp

\label{lastpage}

\end{document}